\documentclass[a4paper,10pt]{article}

\usepackage[a4paper]{geometry}
\usepackage[T1]{fontenc}
\usepackage{textcomp}
\usepackage{newpx}
\usepackage{amsmath}
\usepackage[scaled]{beramono}
\usepackage{pifont}
\usepackage{microtype}

\usepackage{changepage}
\usepackage{layout}
\usepackage{parskip}
\usepackage{epigraph}
\usepackage{fancyvrb}
\usepackage{graphicx}
\usepackage{color}
\usepackage[dvipsnames, table]{xcolor}
\usepackage{listings}
\usepackage{mathpartir}
\usepackage[makeroom]{cancel}
\usepackage{tikz}
\usepackage{etoolbox}
\usepackage{enumitem}
\usepackage{array}
\usepackage{tabularx}
\usepackage{booktabs}
\usepackage[most]{tcolorbox}
\usepackage{float}
\usepackage{subcaption}
\usepackage{multirow}
\usepackage[onehalfspacing]{setspace}
\usepackage{siunitx}

\usepackage[
    backend=bibtex,
    citestyle=numeric,
    sorting=nty,
    bibstyle=ieee,
    maxbibnames=99
]{biblatex}
\DeclareFieldFormat{labelnumberwidth}{\mkbibbrackets{#1}}
\usepackage[colorlinks]{hyperref}
\hypersetup{
    hidelinks,
    linktoc=all,
    bookmarksdepth=4
}

\usepackage{pgfkeys}
\newcommand{\textmonth}{\ifcase \month \or January\or February\or March\or %
April\or May \or June\or July\or August\or September\or October\or November\or %
December\fi}

\newcommand\code[1]             {{\bf\ttfamily #1}}

\definecolor{cobalt}{rgb}{0.0, 0.28, 0.67}

\definecolor{cobaltlight}{HTML}{e8edf8}

\definecolor{SignaloidGreen}{RGB}{48, 125, 43}

\definecolor{plotblue}{HTML}{bfdfff}
\definecolor{plotgreen}{HTML}{caf2db}
\definecolor{plotyellow}{HTML}{fcebcf}

\definecolor{a}{rgb}{0.9,0.95,0.95}
\definecolor{b}{rgb}{0.99,0.99,0.99}

\newcounter{rownumber}[figure]

\newcommand{\circled}[2][]{%
  \tikz[baseline=(char.base)]{%
    \node[shape = circle, draw, inner sep = 1pt]
    (char) {\phantom{\ifblank{#1}{#2}{#1}}};%
    \node at (char.center) {\makebox[0pt][c]{#2}};}}
\robustify{\circled}

\makeatletter
\newcommand{\CurrentNumber}{%
  \ifnum\value{subparagraph}>0
    \thesubparagraph
  \else\ifnum\value{paragraph}>0
    \theparagraph
  \else\ifnum\value{subsubsection}>0
    \thesubsubsection
  \else\ifnum\value{subsection}>0
    \thesubsection
  \else\ifnum\value{section}>0
    \thesection
  \else\ifnum\value{chapter}>0
    \thechapter
  \else
    0%
  \fi\fi\fi\fi\fi\fi
}
\makeatother

\tcbuselibrary{skins, breakable}
\newtcolorbox{focusbox}[1][]{%
  breakable,
  enhanced,
  colback=lightgray!10,
  colframe=blue!40,
  boxrule=0.6pt,
  arc=2mm,
  left=1.5mm,
  right=3mm,
  top=3mm,
  bottom=3mm,
  before skip=1.0\baselineskip,
  after skip=1.0\baselineskip,
  fonttitle=\bfseries,
  title={},
  #1
}

\RenewDocumentEnvironment{focusnotes}{ O{} +b }{}{}

\definecolor{C0}{HTML}{1F77B4} 
\definecolor{C1}{HTML}{FF7F0E} 
\definecolor{C2}{HTML}{2CA02C} 

\definecolor{OmnigraffleLightCayenne}{RGB}{176,30,35}
\definecolor{OmnigraffleLightMocha}{rgb}{0.702, 0.267, 0.0}
\definecolor{OmnigraffleSmokeyFern}{RGB}{33,93,15}

\definecolor{C0}{HTML}{1F77B4} 
\definecolor{C1}{HTML}{FF7F0E} 
\definecolor{C2}{HTML}{2CA02C} 

\definecolor{OmnigraffleLightMocha}{rgb}{0.702, 0.267, 0.0}

\definecolor{babyblue}{rgb}{0.54, 0.81, 0.94}
\definecolor{babyblueeyes}{rgb}{0.63, 0.79, 0.95}
\definecolor{ballblue}{rgb}{0.13, 0.67, 0.8}
\definecolor{asparagus}{rgb}{0.53, 0.66, 0.42}
\definecolor{amethyst}{rgb}{0.6, 0.4, 0.8}

\definecolor{comment}{RGB}{61, 136, 145}
\definecolor{string}{RGB}{255,0,0}  
\definecolor{keyword}{RGB}{75,61,179} 

\definecolor{listinggreen}{rgb}{0,0.6,0}
\definecolor{listinggray}{rgb}{0.5,0.5,0.5}
\definecolor{listingmauve}{rgb}{0.58,0,0.82}
\definecolor{listingkeywordcolor}{rgb}{1.0,0.4,0.0}
\definecolor{listinglightgray}{rgb}{0.98,0.98,0.98}

\lstdefinelanguage{FSharp}
{
	morekeywords	= {
    let,
    type,
    Measure,
	},
	sensitive	= false,
	morecomment	= [l]{\#},
	morecomment	= [s]{(*}{*)},
}

\lstdefinelanguage{Newton}
{
	morekeywords	= {
		signal,
		derivation,
		symbol,
		name,
		invariant,
		constant,
		English,
		sensor,
		name,
		none,
		dot,
		cross,
		derivative,
		integral,
		interface,
		i2c,
		spi,
		analog,
		write,
		read,
		delay,
		range,
		erasuretoken,
		uncertainty,
		accuracy,
		precision,
		Gaussian,
		exponential,
		biexponential,
		to,
		bits,
		dimensionless,
		include
	},
	sensitive	= false,
	morecomment	= [l]{\#},
	morecomment	= [s]{/*}{*/},
}

\lstdefinestyle{c}{
  backgroundcolor=\color{listinglightgray},
	commentstyle=\color{comment},
	stringstyle=\color{string},
	keywordstyle=\color{amethyst},
	basicstyle=\footnotesize\ttfamily,
	numbers=left,
	numberstyle=\tiny,
	numbersep=5pt,
	frame=lines,
	breaklines=true,
	prebreak=\raisebox{0ex}[0ex][0ex]{\ensuremath{\hookleftarrow}},
	showstringspaces=false,
	upquote=true,
	tabsize=2,
	deletekeywords = {double,void},
	keywordstyle = [2]\color{keyword},
	morekeywords = [2]{
    double,
    int,
    size_t,
    uint8_t,
    void},
  emphstyle = [1]\color{ballblue},
  emph      = [1]{
  sample,
  round,
  isnan,
  observation_model_ideal,
  observation_noise_model,
  evaluate_likelihood_analytical,
  evaluate_density_analytical,
  evaluate_likelihood_uxhw,
  evaluate_density_uxhw,
  observation_model,
  observation_model_uxhw,
  transition_model,
  transition_model_ideal,
  transition_noise_model},
  emphstyle = [2]\color{SignaloidGreen},
  emph      = [2]{
  DoubleSample,
  Mixture,
  GaussDist,
  UniformDist,
  LaplaceDist,
  BayesLaplace,
  LimitSupport,
  SupportMin,
  SupportMax,
  IndependentCopy,
  ProbabilityGT,
  Quantile,
  Sample
  },
}

\title{Dynamic Online Processor-Native Inference for State Estimation}
\author{Orestis Kaparounakis}
\date{\today}

\begin{document}
\maketitle

\begin{abstract}
\noindent
Sensor-rich data-driven applications increasingly use Bayesian approaches to
infer latent states of dynamic systems from noisy sensor measurements and
physical models. Yet the computation of the likelihood remains an essential
bottleneck for accurate posteriors and performant inference. This paper presents
a Bayesian filtering technique that uses processor-native uncertainty tracking
for both uncertainty propagation and inference. The technique implements
deterministic hierarchical importance restructuring through a native operation,
giving deterministic latency and bounded memory use for arbitrary models written
as program code. Benchmarks across three nonlinear state-space systems compare
the approach against particle filters and Monte-Carlo-based likelihood
estimators. The technique enables deterministic approximate filtering with as
high as 805\texttimes{} average speedup against direct Monte Carlo work at
matched result quality for model evaluation, and Pareto-dominant
accuracy--latency trade-offs for posterior inference while remaining competitive
in RMSE with baseline particle filters.
\end{abstract}

Sensor-rich data-driven applications increasingly use Bayesian approaches to
infer the latent states of dynamic systems from noisy sensor measurements and
physical models for more reliable state estimation and control. Such systems
rely more and more on nonlinear sensing technologies which, combined with
highly-nonlinear dynamics, produce non-Gaussian posteriors and intractable
likelihoods. The common approach for nonlinear non-Gaussian systems is particle
filtering~\cite{gordon1993novel}. Yet the computation of the likelihood remains
an essential bottleneck for accurate posteriors and performant inference,
especially when the likelihood requires marginalisation over latent noise
sources. 

This paper examines an enhancement to Bayesian filtering on
uncertainty-extended processors that leverages a microarchitectural abstraction
for Bayesian inference. More specifically,
Section~\ref{SectionUxHwBayesLaplaceFilter} contributes a
tunably-approximative approach to Bayesian filtering with arbitrarily complex
state space models. The approach builds on uncertainty-extended processors
which perform deterministic arithmetic and inference on probability density
functions, and enables simple system model formulation as transformations on
random variables in conventional programming language syntax. The UxHw-based
approach has deterministic output and latency, and does not rely on extraneous
random sampling for model simulation, making it ideal for real-time systems.

Section~\ref{SectionInferenceImportanceSampling} revisits importance sampling
which is the inference mechanism most common in particle filtering.
Section~\ref{SectionDeterministicBayesianFiltering} discusses the mechanism for
real-time processor-native deterministic Bayesian inference that underpins the
UxHw-based approach, and relates this mechanism to importance sampling and
deterministic methods for inference.

Section~\ref{SectionBayesLaplaceBenchmarks} experimentally validates the
accuracy of proposed UxHw-based filtering approach compared to the particle
filter for three stochastic system templates. The section also includes
dissected experimental validation of processor-native uncertainty-tracking
accuracy and bare-metal speed of the proposed approach examining the six
constituent models.

\newcommand{\pr}{\operatorname{Pr}}

\section{Bayesian inference with importance sampling}
\label{SectionInferenceImportanceSampling}

Bayesian inference is the mathematical application of Bayes' theorem to infer an
updated \emph{posterior} random variable from a \emph{prior} information random
variable and an observation or \emph{evidence}. Let $X$ denote the prior random
variable with support $\mathcal{X}$, and let $Y$ denote the observation random
variable with realised value $y$. Equation~\ref{EqBayesTheorem} is Bayes'
theorem for continuous random variables:
\begin{equation}
    p_{X \mid Y}(x \mid y)
    =
    \frac{
        p_{Y \mid X}(y \mid x)\, p_X(x)
    }{
        \int_{\mathcal{X}} p_{Y \mid X}(y \mid x')\, p_X(x')\, dx'
    }.
    \label{EqBayesTheorem}
\end{equation}
The theorem states that the posterior density is proportional to the prior
density times the likelihood. Values of $x$ that were plausible under the prior
and that also explain the observation $y$ well receive greater posterior mass.
In many practical systems the normalising integral in the denominator has no
closed form, and even when it does, direct evaluation can be costly in
high-dimensional or nonlinear settings. This difficulty motivates approximation
methods such as importance sampling.

Importance sampling approximates the posterior by sampling a proposal density
which is easier to sample than the posterior itself and then assigning each
sample a corrective weight. Let $q(x \mid y)$ denote a proposal density on
$\mathcal{X}$, and $x^{(i)}$ for $i = 1, \dots, N$ iid samples from the
proposal:
\begin{equation}
    x^{(i)} \sim q(x \mid y),
    \qquad i = 1, \dots, N.
\end{equation}
The corresponding unnormalised importance weights are
\begin{equation}
    \tilde{w}^{(i)}
    =
    \frac{
        p_{Y \mid X}(y \mid x^{(i)})\, p_X(x^{(i)})
    }{
        q(x^{(i)} \mid y)
    },
    \qquad i = 1, \dots, N.
\end{equation}
The normalised weights are
\begin{equation}
    w^{(i)}
    =
    \frac{\tilde{w}^{(i)}}{\sum_{j=1}^{N} \tilde{w}^{(j)}},
    \qquad i = 1, \dots, N.
\end{equation}
These weighted samples form a discrete approximation of the posterior. This is
the idea that underpins the particle filter, where the proposal generates
particles at each time step and the importance weights correct for the
difference between the proposal and the target filtering distribution.
Resampling the weighted set yields an unweighted set of posterior samples.

For importance sampling to be valid, the proposal must assign positive density
wherever the unnormalised posterior $p_{Y \mid X}(y \mid x)\, p_X(x)$ is
positive. In other words, the support of $q(x \mid y)$ must contain the support
of the posterior. If this condition does not hold, the process never samples
these regions of the posterior mass and the approximation becomes biased. In
practice, engineers choose proposals that reasonably resemble the posterior,
because a poor mismatch produces uneven weights and results in low-quality
approximation~\cite{doucet2000sequential}.

As $N \to \infty$, for a well-defined posterior and given that the proposal
assigns positive density wherever the posterior has positive density, the
weighted empirical distribution $\{x^{(i)}, w^{(i)}\}_{i=1}^N$ converges to the
true posterior distribution.

\paragraph*{Other posterior-approximation methods}
Markov-chain Monte Carlo methods approximate the posterior by constructing a
Markov chain whose stationary distribution is the posterior
\cite{metropolis1953equation,hastings1970monte}. Variational inference
approximates the posterior by solving an optimisation problem over a tractable
family of distributions \cite{jordan1999introduction,blei2017variational}.
Laplace approximation replaces the posterior locally with a Gaussian
approximation around a mode \cite{tierney1986accurate}.

\section{Real-time processor-native deterministic Bayesian inference}
\label{SectionDeterministicBayesianFiltering}

Processors capable of native uncertainty tracking may also include native
support for native probabilistic inference. The UxHw system includes the
BayesLaplace operation: a native processor operation for Bayesian
inference~\cite{bilgin2025bayeslaplace,tsoutsouras2021laplace} which implements
\emph{deterministic importance restructuring}.

Inference algorithms using the UxHw system use the same programming interface
for both the inference program and the modelling code. This contrasts
probabilistic programming languages which typically have different syntaxes for
modelling code and inference code~\cite{cusumano2019gen}.

\paragraph*{Gradual uncertainty specification} Implementing Bayesian filtering
with processor-native uncertainty tracking enables the specification of
uncertainty for more parameters than conventionally typical for other filtering
methods. During different stages of the design, engineers can tune the
distribution specifications and also choose to model more program variables as
uncertain variables. For example, using the UxHw system for filtering
trivialises modelling the uncertainty of the transition model input (control
input), and other model parameters, without having to readjust the whole filter.

\paragraph*{Multi-model estimation at region boundaries} Some estimation
problems use different models in different regions of the state space. When the
active model depends on the current state, an implementation can branch on
uncertain state values and carry the resulting hypotheses jointly instead of
using a point statistic to select a single model. This paper does not benchmark
multi-model estimation directly. The same idea
also appears in systems such as sensors that switch between different
approximations across operating regions.

\paragraph*{Empirical likelihoods} Some Bayesian filtering problems lack
an observation model in either analytic or programmatic form. In such cases, engineers may
instead use prior labelled observations to build an empirical likelihood proxy.
Processor-native uncertainty tracking makes this style of implementation a
natural fit, because the filter can evaluate such proxies directly in code and
combine them with uncertain state representations without requiring a
parametric observation model. This paper does not evaluate empirical
likelihoods directly, but they are another example of how the same programming
model could support filtering designs beyond the standard closed-form setting.

\subsection{Deterministic hierarchical importance restructuring}
\label{SectionDeterministicHierarchicalRestructuring}

The particle filter algorithm uses weighted resampling which performs stochastic
ancestor selection based on the particle
weights~\cite{gordon1993novel,johansen2009tutorial}. In contrast, in the
BayesLaplace operation of the UxHw system, the representation restructuring step
deterministically transforms a weighted empirical measure into another weighted
empirical measure of fixed size using both the particle weights and state
values~\cite{bilgin2026quantization}. This algorithm is conceptually close to
importance sampling with the following distinctions:
\begin{itemize}
    \item Determinism: The algorithm does not depend on random sampling. The
    same prior, likelihood, and evidence always yield the same posterior.
    \item Importance: The underling UxHw representation enables the use of any
    sampling model as a likelihood function without needing to analytically
    derive a density function.
    \item Restructuring: The samples of the posterior adhere to a
    weight-position spacing rule~\cite{bilgin2026quantization}. The posterior
    positions do not only depend on the weight as in conventional importance
    sampling.
\end{itemize}

In a moment-matching manner, the UxHw representation restructures the weighted
particles using a recursive algorithm that splits the domain into regions of
equal particle mass~\cite{bilgin2025quantization}. This approach resembles deterministic self-fission
resampling~\cite{li2012deterministic} with the difference that UxHw does not
require heuristic-dependent hyperparameters such as the starting position and
density threshold. Instead of aiming for final positions of similar mass, UxHw
splits existing bins in two and aims for a specific count $\eta$ of final
positions. This $\eta$ is a configuration parameter of UxHw controlling the
processor speed versus distribution arithmetic fidelity trade-off.

Let $\mathtt{x}$ denote a UxHw program variable that encodes the prior, and let $X
\coloneqq \operatorname{rv}(\mathtt{x})$ denote the corresponding ideal random
variable with law $\mu_X$. Let $\eta$ denote the UxHw size configuration
parameter. The UxHw system represents $\mathtt{x}$ by a weighted set of support
points $\{x^{(i)}, w^{(i)}\}_{i=1}^\eta$ that satisfies the telescoping-torques
structural condition~\cite{bilgin2026quantization}. Let
\begin{equation}
    \hat{\mu}_{\mathtt{x}}
    \coloneqq
    \sum_{i=1}^{\eta} w^{(i)} \delta_{x^{(i)}}.
\end{equation}
Then $\hat{\mu}_{\mathtt{x}}$ is the implementation-level law encoded by
$\mathtt{x}$:
\begin{equation}
    \hat{\mu}_{\mathtt{x}} \approx \mu_X.
    \label{EqImportanceRestructuringUxHwRepresentation}
\end{equation}

Let $h$ denote the program function that encodes the observation model: for a
given hypothesis $x$, $h(x)$ is the distribution of probable observations. Let
$z$ denote a realised observation. The BayesLaplace operation computes the
posterior program variable $\mathtt{x}^{+}$~\cite{bilgin2025bayeslaplace}:
\begin{equation}
    \mathtt{x}^{+}
    \coloneqq
    \operatorname{BayesLaplace}(h, \mathtt{x}, z, 1).
    \label{EqImportanceRestructuringBayes}
\end{equation}
Let $X^{+} \coloneqq \operatorname{rv}(\mathtt{x}^{+})$ denote the corresponding
ideal posterior random variable with law $\mu_{X^{+}}$, and let
\begin{equation}
    \hat{\mu}_{\mathtt{x}^{+}}
    \coloneqq
    \sum_{i=1}^{\eta} \hat{w}^{(i)} \delta_{\hat{x}^{(i)}}
\end{equation}
denote the law encoded by the UxHw representation
$\{\hat{x}^{(i)}, \hat{w}^{(i)}\}_{i=1}^{\eta}$. Then
\begin{equation}
    \hat{\mu}_{\mathtt{x}^{+}} \approx \mu_{X^{+}}.
\end{equation}

The following steps summarise the deterministic hierarchical restructuring that
the BayesLaplace operation performs~\cite{bilgin2025bayeslaplace}:
\begin{enumerate}
    \item For each prior support point $x^{(i)}$, compute the corresponding
    observation-predictive program variable
    \begin{equation}
        \mathtt{y}^{(i)} \coloneqq h(x^{(i)}),
        \qquad i = 1, \dots, \eta.
    \end{equation}
    Let $Y^{(i)} \coloneqq \operatorname{rv}(\mathtt{y}^{(i)})$ denote the
    corresponding ideal observation random variable with law $\mu_{Y^{(i)}}$,
    and let $\hat{\mu}_{\mathtt{y}^{(i)}}$ denote the law encoded by
    $\mathtt{y}^{(i)}$. Each $Y^{(i)}$ is conditioned on $x^{(i)}$. Each
    $\mathtt{y}^{(i)}$ has a representation inside the UxHw system of the
    following form:
    \begin{equation}
        \hat{\mu}_{\mathtt{y}^{(i)}}
        \coloneqq
        \sum_{j=1}^{\eta} \omega^{(i,j)} \delta_{y^{(i,j)}},\qquad i = 1, \dots, \eta.
    \end{equation}
    This encoded law approximates the ideal conditional observation law:
    \begin{equation}
        \hat{\mu}_{\mathtt{y}^{(i)}} \approx \mu_{Y^{(i)}},\qquad i = 1, \dots, \eta.
    \end{equation}
    At this stage, the collection of conditional observation laws forms a Dirac
    mixture, and does not satisfy the structural constraints of the UxHw
    representation~\cite{bilgin2026quantization}.

    \item Update the prior weights using the likelihood of the realised
    observation $z$:
    \begin{equation}
        \tilde{w}^{(i)}
        \coloneqq
        w^{(i)}\,
        \operatorname{EvaluatePDF^\star}\!\big(\mathtt{y}^{(i)}, z\big),
        \qquad i = 1, \dots, \eta. \label{EqDhirWeightUpdate}
    \end{equation}
    Here, $\operatorname{EvaluatePDF^\star}$ denotes the implementation-level
    density-evaluation logic of $\operatorname{EvaluatePDF}$
    as it operates internally on intermediate variables represented by Dirac
    mixtures.

    \item Normalise the updated weights:
    \begin{equation}
        \tilde{w}^{(i)}
        \leftarrow
        \frac{\tilde{w}^{(i)}}{\sum_{k=1}^{\eta} \tilde{w}^{(k)}},
        \qquad i = 1, \dots, \eta.
    \end{equation}

    \item This yields the deterministic-importance posterior, before
    restructuring, with encoded law of Equation~\ref{EqDhirPosteriorLaw}:
    \begin{equation}
        \tilde{\mu}_{\mathtt{x}^{+}}
        \coloneqq
        \sum_{i=1}^{\eta} \tilde{w}^{(i)} \delta_{x^{(i)}}. \label{EqDhirPosteriorLaw}
    \end{equation}
    Equivalently, the support-point representation is
    $\{x^{(i)}, \tilde{w}^{(i)}\}_{i=1}^{\eta}$.

    \item Apply the telescoping-torques restructuring algorithm to enforce the
    structural constraints of the UxHw
    representation~\cite{bilgin2026quantization}:
    \begin{equation}
        \{\hat{x}^{(i)}, \hat{w}^{(i)}\}_{i=1}^{\eta}
        \coloneqq
        \operatorname{TTR}\!\big(\{x^{(i)}, \tilde{w}^{(i)}\}_{i=1}^{\eta}\big).
        \label{EqDhirPosteriorStructured}
    \end{equation}
    Equation~\ref{EqDhirPosteriorLawStructured} is the law encoded by the
    posterior program variable $\mathtt{x}^{+}$:
    \begin{equation}
        \hat{\mu}_{\mathtt{x}^{+}}
        \coloneqq
        \sum_{i=1}^{\eta} \hat{w}^{(i)} \delta_{\hat{x}^{(i)}}. \label{EqDhirPosteriorLawStructured}
    \end{equation}
\end{enumerate}

\subsection{Empirical observation densities}
\label{SectionInferenceEmpiricalObservationDensities}

The UxHw semantics of Bayesian inference allow the observation input to be an
empirical observation law rather than a single realised value. This is useful
when sensors sample at a higher frequency than the filter update rate, where
conventional implementations often replace a measurement bundle by its average.
Using an empirical observation law instead reduces modelling assumptions.

Let the observation program variable $\mathtt{z}$ encode an empirical observation
law. Let $\hat{\mu}_{\mathtt{z}}$ denote the corresponding encoded law with
support-point representation $\hat{\mu}_{\mathtt{z}} \coloneqq \sum_{j=1}^{\eta}
\omega^{(j)} \delta_{z^{(j)}}$.

Let $m \in \mathbb{N}$ the count of iid observations from which $\mathtt{z}$ was
created. The BayesLaplace
operation also needs $m$ to know the amount of independent observations that $\mathtt{z}$ represents.
Equation~\ref{EqDhirBayesLaplaceMultiObservation} shows the application of the
BayesLaplace operation in this multi-observation distribution scenario:
\begin{equation}
    \mathtt{x}^{+}
    \coloneqq
    \operatorname{BayesLaplace}(h, \mathtt{x}, \mathtt{z}, m).
    \label{EqDhirBayesLaplaceMultiObservation}
\end{equation}

For this multi-observation case, the weight update of
Equation~\ref{EqDhirWeightUpdate} scales the encoded empirical law to represent
a bundle of size $m$. At the representation level, this gives
\begin{equation}
    \tilde{w}^{(i)}
    \coloneqq
    w^{(i)}
    \prod_{j=1}^{\eta}
    \left(
        \operatorname{EvaluatePDF^\star}\!\big(\mathtt{y}^{(i)}, z^{(j)}\big)
    \right)^{m \omega^{(j)}},
    \qquad i = 1, \dots, \eta.
    \label{EqDhirWeightUpdateMultiObservation}
\end{equation}
Equation~\ref{EqDhirWeightUpdateMultiObservation} expresses the fact that
$\mathtt{z}$ encodes an empirical distribution with $\eta$ support points, while the
parameter $m$ scales that distribution so that it contributes the joint
likelihood of $m$ independent observations.

The empirical observation density approach is attractive in situations where
sensors generate samples in a frequency higher than what the system estimator
wants. Conventionally, in such cases, the algorithm implementation averages the
a sequence of samples before passing them to the estimator.

\section{The UxHw BayesLaplace filter}
\label{SectionUxHwBayesLaplaceFilter}

\begin{figure}
\centering
\captionsetup{type=figure}
\includegraphics[trim={0cm 0cm 0cm 0cm},clip,width=0.99\textwidth]{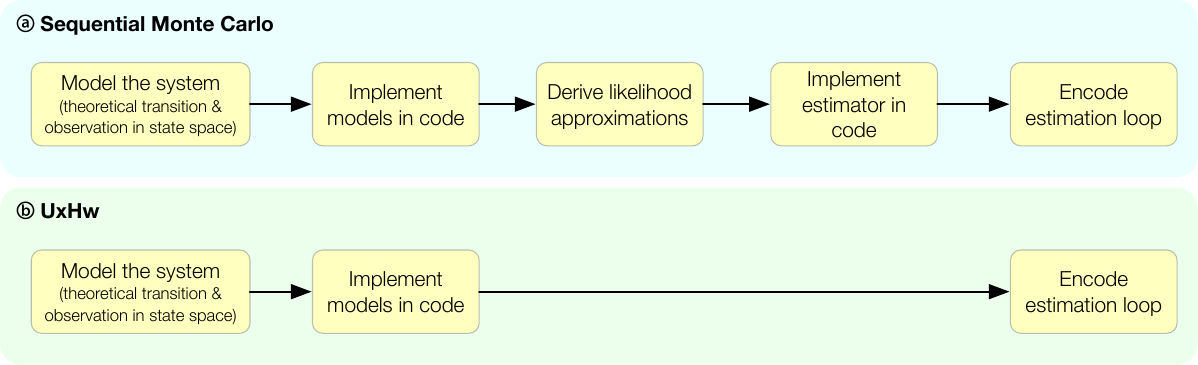}
\caption[High-level view of the engineering process for implementing state
estimation with sequential Monte Carlo and with processor-native deterministic
uncertainty tracking.]{High-level view of the engineering process for
implementing state estimation \textbf{\textcircled{\small a}} with sequential
Monte Carlo and \textbf{\textcircled{\small b}} with UxHw. In sequential Monte
Carlo approaches engineers must derive the likelihood function either by
analytical derivation when possible or by computational approximation, and then
must write the code for the inference mechanism that uses it. Conversely, in the
UxHw approach the model code doubles as the likelihood function and the UxHw
system provides the inference mechanism as a native operation.}
\label{FigureBayesianFilteringDesignProcess}
\end{figure}

This section introduces a Bayesian filtering technique which enables
highly expressive filter implementations while also offering better posteriors,
increased performance, and avoiding extraneous random number generation. This
approach builds on recent advances in hardware architectures for performing
arithmetic on digital representations of probability density
functions~\cite{tsoutsouras2022laplace, bilgin2026quantization}. This section
proposes a Bayesian filtering method which leverages processor-native Bayesian
inference~\cite{bilgin2025bayeslaplace} provided by the UxHw system.

Figure~\ref{FigureBayesianFilteringDesignProcess} shows a high-level view of the
typical engineering process for implementing state estimation for a stochastic system:
\textbf{\textcircled{\small a}} for the conventional approach (particle
filtering~\cite{gordon1993novel,pitt1999filtering} and Kalman
filtering~\cite{kalman1961new}), and \textbf{\textcircled{\small b}} for the
UxHw-based approach. In the conventional approach, after modelling the system
and implementing it in code, engineers must derive the likelihood function
either by analytical derivation when possible or by computational approximation,
and then must write the code for the inference mechanism that uses it.
Conversely, in the UxHw approach the model code doubles as the likelihood
function and UxHw provides the inference mechanism as a native operation and no
further work necessary other than the outer estimation loop.

Let $\mathbf{x}_{k}$ represent the system state at time step $k$. In the state
transition (Equation~\ref{eq:NonlinearProcess}), the function $f$ models the
state transition dynamics and takes into account the system input vector
$\mathbf{u}_{k}$. The vector $\mathbf{w}_{k}$ represents latent system
perturbations (\emph{process noise}). In the measurement
(Equation~\ref{eq:NonlinearMeasurement}), function $h$ models the conversion of
the measurand to the measurement, and the vector $\mathbf{v}_{k}$ represents
latent \emph{measurement noise}.
\begin{align}
    \mathbf{x}_{k} &= f(\mathbf{x}_{k-1}, \mathbf{u}_{k}) + \mathbf{w}_{k}, \label{eq:NonlinearProcess} \\
    \mathbf{z}_{k} &= h(\mathbf{x}_{k}) + \mathbf{v}_{k}. \label{eq:NonlinearMeasurement}
\end{align}

Let $L\langle\cdot,\cdot\rangle$ represent computation on hardware with native
uncertainty tracking. Let $\operatorname{BL}$ represent processor-native
Bayesian inference. Equation~\ref{EqBayesLaplacePredict}
and~\ref{EqBayesLaplaceUpdate} summarise the form of Bayesian inference in this
setting:
\begin{align}
    \mathnormal{\textbf{Predict:}}\quad& \hat{\mathbf{x}}_{k|k-1} = L\langle f, (\hat{\mathbf{x}}_{k-1|k-1}, \mathbf{u}_{k})\rangle \label{EqBayesLaplacePredict}\\
    \mathnormal{\textbf{Update:}}\quad& \hat{\mathbf{x}}_{k|k} = \operatorname{BL}\hspace{-0.4ex}\left( L\langle h , \hat{\mathbf{x}}_{k|k-1}\rangle, \mathbf{z}_{k}\right) \label{EqBayesLaplaceUpdate}
\end{align}
The action $L$ lifts endomorphisms from the real domain to the domain of
probability measures on the real line and applies the argument.

\begin{figure}[htb]
\centering
\begin{lstlisting}[language=C,style=c,caption={[1D UxHw BayesLaplace filter main
loop source code implementation in C.]Source code in C for the 1D UxHw 
BayesLaplace filter.
In the BayesLaplace filter, the transition and observation models are both C 
functions that implement a model sampler which, when run on UxHw, produces the 
predictive distribution of each model conditional on the input uncertainties. To
improve memory usage, this implementation uses the IndependentCopy operation
to decorrelate program variables no longer necessary.},label=ListingUxHwBayesLaplaceFilterCode]
double  observation_model(void *  optional_extra_args, double point_hypothesis);
double  transition_model(double previous_state, double transition_noise, /* ...extra arguments */);

/* Initial state estimate as Gaussian(x_0, sqrt(init_variance)) */
predicted_state[0] = GaussDist(true_state[0], sqrt(init_variance));

/* UxHw BayesLaplace filter main loop */
for (size_t t = 0; t < state_history_len; ++t) {
    double predictive_state_dist, updated_state;
    if (t > 0) {  /* The first observation is of the initial state */

        predictive_state_dist = transition_model(
                                    estimated_state[t-1],
                                    transition_noise_model(),
                                    /* ... any extra arguments */
                                );

        predicted_state[t] = IndependentCopy(predictive_state_dist);  /* Memory conservation */
    }
    
    updated_state = BayesLaplace(
                        observation_model,      /* Function pointer to the observation model */
                        &optional_extra_args,
                        predicted_state[t],     /* Prior distribution */
                        observed_state[t],      /* Observed data (point value or distribution) */
                        1                       /* Observations count */
                    );
    if (isnan(updated_state)) {
        updated_state = predicted_state[t];
    }

    estimated_state[t] = IndependentCopy(updated_state);    /* Memory conservation */
}
\end{lstlisting}
\end{figure}

\subsection{1D BayesLaplace filter implementation in C for the UxHw system}

Listing~\ref{ListingUxHwBayesLaplaceFilterCode} gives source code in C for a
generic 1D implementation of the processor-native-inference BayesLaplace filter
of Equations~\ref{EqBayesLaplacePredict} and~\ref{EqBayesLaplaceUpdate}
targeting the UxHw system. In the framework of the BayesLaplace filter both the
transition and observation models are functions that resemble samplers for the
respective model. When running on UxHw, the noise models produce fresh variables
with associated uncertainty information, and the system model functions yield
predictive distributions of the respective quantity. I.e.,
\code{transition\_model} yields the state-predictive distribution and
\code{observation\_model} yields the observation-predictive distribution
conditioned on the passed \code{point\_hypothesis} value. The state-predictive
distribution serves as the prior in the BayesLaplace operation, and the
microarchitectural implementation breaks it into point hypotheses which it
passes into the observation model. This process produces updated weights for the
prior model based also on the likelihood of the observation data which
ultimately yields the posterior distribution as the output of the BayesLaplace
operation.

This implementation includes simple guardrails for improved memory usage and
correctness. The implementation uses the IndependentCopy operation to
decorrelate program variables no longer necessary. Applying this operation lines up
with the expected Markov property of systems this works examines. For processes
with more memory, the filter would need a mechanism to apply the operation at
the memory horizon variables, and at the same time to leave all variables within
the filter memory properly correlated. 

The BayesLaplace operation adheres to the dogmatic property of Bayesian
inference whereby areas of the domain where the prior assigns zero density
cannot have non-zero density in the posterior. In other words, Bayes' rule cannot create
belief from nothing even if there is evidence. This makes it important for the
prior to accurately capture the underlying distribution of possible states. In
the context of UxHw, distribution representations have finite support which can
create a problem for extreme outlier observations---although to smaller extend
than random-sampling approaches such as the particle filter with bounded observation models.
To address this error mode, the
implementation checks the result of the BayesLaplace operation and, on detected
failure, sets the posterior from the prior to keep the filter running.

\begin{figure}
    \centering
    \includegraphics[trim={0cm 0cm 0cm 0cm},clip,width=0.75\columnwidth]{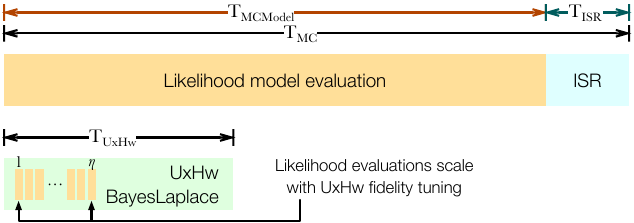}
    \caption[Illustrative timing diagram of importance sampling and
    deterministic hierarchical importance restructuring.]{Illustrative timing
    diagram of importance sampling and deterministic hierarchical importance
    restructuring via BayesLaplace. To perform the weighting step, the particle
    filter must compute the likelihood of the observation for every particle.
    This operation scales with the number of particles and can become expensive
    for intractable-likelihoods which also require marginalisation over noise
    sources. With processor-native uncertainty tracking, UxHw is able to glean
    more information from computations which scale with the UxHw representation
    size.\label{FigureBeerLambertInferenceTiming}}
\end{figure}

Figure~\ref{FigureBeerLambertInferenceTiming} shows a illustrative timing
diagram for a single inference step using the Monte Carlo method (importance
sampling) and using UxHw~8. To perform the weighting step, the particle filter
must compute the likelihood of the observation for every particle. This
operation scales with the number of particles and can become expensive for
intractable-likelihoods which also require marginalisation over more than one noise sources.
When the particles count is
in the hundreds or low thousands, the likelihood model dominates the compute
budget of the particle filter. Conversely, the UxHw approach computes
observation-predictive distributions whose count scales with the UxHw
representation size. With processor-native uncertainty tracking, UxHw is able to
glean more information from fewer points at expedient rate.

\section{Benchmarks}
\label{SectionBayesLaplaceBenchmarks}

This paper benchmarks the UxHw approach on three nonlinear systems stemming
from practical application scenarios. The benchmarks test the filtering accuracy
of proposed UxHw-based approach compared to the particle filter for three
stochastic system templates across 150 traces. Each section also includes
dissected experimental validation of uncertainty-quantification accuracy and
bare-metal speed of the proposed approach examining the six constituent
transition models and observation models running on a commercially-available
uncertainty-tracking system-on-module across 273 distinct configurations. They also include
experimental validation of the posterior-accuracy and bare-metal speed of the
UxHw inference mechanism against the importance sampling of particle filters.
Table~\ref{TabBenchmarkSystems} lists the system configurations with which this
paper benchmarks the UxHw-based BayesLaplace filtering approach.

Because the UxHw-based filters shown here and standard particle filters both
evaluate the observation model more times than the transition model, because of
the likelihood computation, the observation model has more impact in the overall
accuracy and latency of the filter compared to the transition model. (As a
counter-example, the auxiliary particle filter additionally evaluates the
transition model when computing the likelihood.)

\begin{table}[t]
\centering
\caption[Summary of systems under test and constituent models for Bayesian
filtering.]{Summary of systems under test and constituent models for Bayesian
filtering using the UxHw system. The systems all have additive transition uncertainty
specification while the observation uncertainty differs.}
\footnotesize
\begin{tabular}{r r|l l}
\hline
 Transition noise & \textbf{Transition model} &  \textbf{Observation model} & Observation noise \\
\hline
\rowcolor{a} Additive & \multicolumn{2}{c}{Stochastic volatility system (Eq.~\ref{EqStochasticVolatilityProcess},\,\ref{EqStochasticVolatilityObservation})} & Multiplicative \\
\rowcolor{b} Additive & \multicolumn{2}{c}{Gordon--Salmond--Smith system (Eq.~\ref{EqGordonWalkTransitionBayesLaplace},\,\ref{EqGordonWalkObservationBayesLaplace})} & Additive \\
\rowcolor{a} Additive & Cubic mean-reverting (Eq.~\ref{EqCubicRevertingProcess}) & Beer--Lambert (Eq.~\ref{EqBeerLambertObservationModel}) & Multiplicative and heteroscedastic  \\
\hline
\end{tabular}
\label{TabBenchmarkSystems}
\end{table}

In the following sections, $x_k$ denotes the system state at time index $k$, and
$y_k$ the observation.

\subsection{Stochastic volatility system}

A discrete \emph{stochastic-volatility model} links a hidden log-volatility
process to observations whose noise scale changes over
time~\cite{pitt1999filtering, jacquier2004bayesian}. Let $x_k$ be the latent
log-variance at time step $k$, and $y_k$ be the observed return. Let $c \in
\mathbb{R}$ be the drift, and let $\sigma_\eta > 0$ represent the volatility of
log-volatility, and $\phi \in (-1,1)$ the persistence. Also, let the shock term
$\eta_k$ with law $\mu_\eta(\sigma_\eta)$ and $\varepsilon_k$ each denote iid
noise. Equations~\ref{EqStochasticVolatilityProcess}
and~\ref{EqStochasticVolatilityObservation} are the standard stochastic
volatility model:
\begin{gather}
    x_k = \phi x_{t-1} + c (1 - \phi) + \eta_k, \label{EqStochasticVolatilityProcess}\\
    y_k = \varepsilon_k\,e^{0.5 x_k}.   \label{EqStochasticVolatilityObservation}
\end{gather}

This benchmark tests the conventional configuration of the stochastic volatility
system with $\eta_k \sim \mathcal{N}(0, 0.2^2)$, $\varepsilon_k \sim
\mathcal{N}(0, 1^2)$, $c = 0$, $\phi=0.98$ using the UxHw BayesLaplace filter of
Section~\ref{SectionUxHwBayesLaplaceFilter} for configuration sizes $\eta \in
\{8, 16, 32, 64\}$ against an assortment of particle filters with particle
counts $n \in \{10, 20, 50, 100, 500, 1000\}$. The benchmark generates 50 traces
of the stochastic volatility system and each filter runs on each trace 100
times. The benchmark computes the average root mean square error (RMSE) across
all runs and traces as a metric of estimation error, the average per-trace RMSE
standard deviation as a metric of estimator stability, and the negative
log-likelihood score (NLL) and the continuous rank probability score (CRPS) as
posterior quality metrics.

\begin{table}
\centering
\caption[Stochastic volatility system filter accuracy results.]{Estimation accuracy
of different configuration sizes of the UxHw BayesLaplace filter and the
bootstrap particle filter for the stochastic volatility system of
Equations~\ref{EqStochasticVolatilityProcess}
and~\ref{EqStochasticVolatilityObservation} with noise distributions $\eta_k
\sim \mathcal{N}(0, 0.2^2)$, $\varepsilon_k \sim \mathcal{N}(0, 1^2)$, and
initial estimate $x_0 \sim \mathcal{N}(0, 2.25^2)$.}
\label{TableStochasticVolatilityFilterAccuracy}
\footnotesize
\begin{tabular}{lr|S[table-format=1.6]S[table-format=1.6]S[table-format=3.6]S[table-format=1.6]}
\toprule
\textbf{Method} & \textbf{\#Traces} & \textbf{RMSE}\enspace\scalebox{0.8}{\rotatebox{90}{\ding{222}}} & \textbf{Per-trace RMSE Std} & \textbf{NLL} & \textbf{CRPS} \\
\midrule
\rowcolor{a} UxHw 64  &       50 & 0.495185 &           0.000000 &  0.743210 & 0.282312 \\
\rowcolor{b} UxHw 32  &       50 & 0.495827 &           0.000000 &  0.763453 & 0.283125 \\
\rowcolor{a} BPF 1000 &       50 & 0.496254 &           0.010016 &  0.728420 & 0.283270 \\
\rowcolor{b} BPF 500  &       50 & 0.498984 &           0.014842 &  0.738632 & 0.285158 \\
\rowcolor{a} UxHw 16  &       50 & 0.499267 &           0.000000 &  0.819344 & 0.286394 \\
\rowcolor{b} BPF 100  &       50 & 0.511004 &           0.032625 &  0.815526 & 0.294365 \\
\rowcolor{a} UxHw 8   &       50 & 0.526074 &           0.000000 &  0.902752 & 0.306625 \\
\rowcolor{b} BPF 50   &       50 & 0.526232 &           0.045955 &  0.962281 & 0.306609 \\
\rowcolor{a} BPF 20   &       50 & 0.572537 &           0.077646 &  1.691358 & 0.343519 \\
\rowcolor{b} BPF 10   &       50 & 0.645772 &           0.115810 &  4.467825 & 0.404854 \\
\bottomrule
\end{tabular}
\end{table}

\begin{figure}[tb]
    \centering
    \includegraphics[trim={0.1cm 0.1cm 0.1cm 0cm},clip,width=0.9\textwidth]{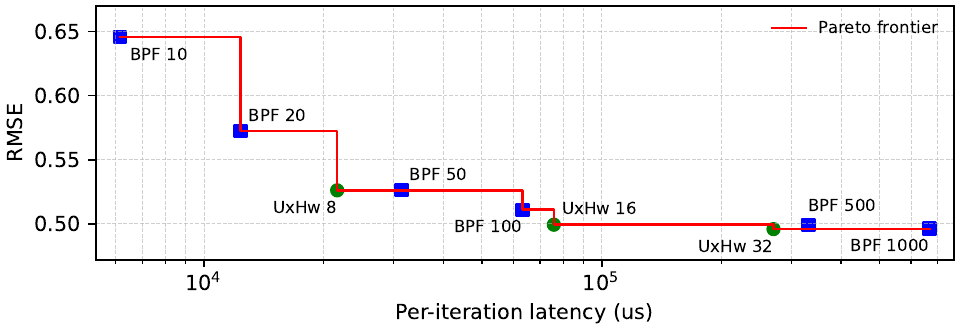}
    \caption[Pareto frontier plot for filtering RMSE and latency of the
    stochastic volatility system.]{Pareto frontier plot for filtering RMSE and
    latency of the stochastic volatility system. Green circles are the
    BayesLaplace filter and blue squares are the bootstrap particle filter. The
    UxHw-based filters achieve dominating positions in the pareto frontier.
    Particle filters beyond 100 particles converge to a better RMSE--latency
    coefficient and the 1000-particle filter dominates at latency
    \qty{270}{\milli\second}.\label{FigStochasticVolatilityRMSELatencyPareto}}
\end{figure}

Table~\ref{TableStochasticVolatilityFilterAccuracy} presents the filtering
benchmark results sorted by ascending RMSE. The lowest estimation error comes
from UxHw~64, which achieves a cross-trace cross-run average RMSE of
\num{0.495185}. UxHw~32 follows with RMSE \num{0.495827}, and the 1000-particle
bootstrap filter reaches \num{0.496254}. The 500-particle bootstrap filter and
UxHw~16 follow closely with \num{0.498984} and \num{0.499267}, respectively.
UxHw~8 reaches \num{0.526074}, which places it essentially alongside BPF~50 at
\num{0.526232}. As expected, the particle filter improves monotonically with
particle count, and the returns diminish at the largest sizes. The UxHw filters
also improve monotonically with configuration size in this benchmark, with
UxHw~64 giving the strongest point-estimate accuracy.
Figure~\ref{FigStochasticVolatilityRMSELatencyPareto} shows that the UxHw-based
filters occupy dominating positions on the RMSE--latency frontier, while the
larger particle filters beyond 100 particles converge to a stronger
RMSE--latency coefficient, with BPF~1000 dominating at a latency of
\qty{270}{\milli\second}.

The per-trace RMSE standard deviation indicates estimation stability. The UxHw
filters rely on deterministic distributional arithmetic, so repeated runs on the
same trace yield identical estimates up to numerical precision in the order of
\num{e-17}. In contrast, the particle filters exhibit run-to-run variability
from stochastic propagation and resampling. This variability increases as the
particle count decreases, with per-trace RMSE standard deviation rising from
\num{0.010016} at 1000 particles to \num{0.115810} at 10 particles. This
contrast makes the UxHw filters attractive when repeatability is an important
requirement.

The negative log-likelihood score measures the probability density that the
posterior assigns to the true latent state, and lower is better. BPF~1000
achieves the best NLL of \num{0.728420}, followed by BPF~500 at
\num{0.738632}. UxHw~64 and UxHw~32 follow closely with \num{0.743210} and
\num{0.763453}, respectively. Among the remaining methods, BPF~100 reaches
\num{0.815526}, UxHw~16 reaches \num{0.819344}, UxHw~8 reaches \num{0.902752},
and BPF~50 reaches \num{0.962281}. The lower-particle bootstrap filters then
degrade more clearly, reaching \num{1.691358} at 20 particles and
\num{4.467825} at 10 particles. For NLL, both method families improve
monotonically with configuration size, and the strongest particle-filter
configurations remain slightly ahead of the strongest UxHw configurations.

The continuous ranked probability score measures calibration and sharpness
around the true latent state, penalising over-spread posteriors, and lower is
better. UxHw~64 achieves the best CRPS of \num{0.282312}, followed by UxHw~32 at
\num{0.283125}. BPF~1000 follows closely with \num{0.283270}, and BPF~500
reaches \num{0.285158}. UxHw~16 remains competitive at \num{0.286394}. Among the
lower-fidelity methods, BPF~100 reaches \num{0.294365}, while BPF~50 and UxHw~8
are nearly identical at \num{0.306609} and \num{0.306625}, respectively. The
ranking by CRPS broadly tracks the ranking by RMSE for the strongest methods.
Overall, UxHw~64, UxHw~32, and BPF~1000 are the strongest methods in this
benchmark, with the UxHw filters giving the best accuracy at the lowest
latencies, and BPF~1000 providing the strongest high-latency point on the
RMSE--latency frontier.

The following subsections dive deeper into the constituent parts of the
BayesLaplace filter: the transition model, the observation model, and the
inference mechanism, and make side-by-side benchmarking with the corresponding
parts of the particle filter algorithm.

\subsubsection{Stochastic volatility transition model benchmark}

The dissected model benchmark tests the speed and quality of UxHw uncertainty
tracking for the stochastic volatility transition model
(Equation~\ref{EqStochasticVolatilityProcess}) on the UxHw-FPGA-17k hardware
module. Let $X_k$ denote the state uncertainty at step $k$, and let $H_k$ denote
the random variable of the shock term $\eta_k$. The benchmark tests for the
usual Gaussian shock and also for a heavier-tailed shock~\cite{asai2009bayesian,
jacquier2004bayesian}, modelled here as an additive Laplace-distributed term.
Let $(\lambda_x, s_x)$ denote the location and scale of $X_k$. Let $\mu$ denote
the parametric family used for $H_k$. Let $H_k$ denote the random variable of
the additive noise. The benchmark tests accuracy of the uncertainty propagation
result for nine hand-picked high-likelihood state locations $\lambda_x$.
Equations~\ref{EqMeanRevertingBenchmarkXkDistribution}--\ref{EqMeanRevertingBenchmarkFixedParameters}
summarise this benchmark configuration space:
\begin{flalign}
    X_k &\sim \mathcal{N}(\lambda_x, s_x), \label{EqMeanRevertingBenchmarkXkDistribution} \\
    \lambda_x &\in \{-2,\,-1,\,-0.5,\,-0.25,\,0,\,0.25,\,0.5,\,1,\,2\}, \text{ and} \label{EqMeanRevertingBenchmarkLambdaValues}\\
    H_k &\sim \mu(0,\sigma_\eta),\qquad \mu \in \{\mathcal{N}, \mathcal{L}(b)\}, \text{ with} \label{EqMeanRevertingBenchmarkNoiseDistribution}\\
    b &= \sigma_\eta / \sqrt{2} \text{ when } \mu = \mathcal{L} \text{ so that } \operatorname{Var}(H_k)=\sigma_\eta^2, \label{EqMeanRevertingBenchmarkLaplaceScale}\\
    s_x &\in \left\{1,\, s_0\right\},\enspace \text{with}\enspace
    s_0 \coloneqq \frac{\sigma_\eta}{\sqrt{1-\phi^2}}, \text{ and} \label{EqMeanRevertingBenchmarkVarianceChoices}\\
    c &= 0,\quad \phi = 0.98,\quad \sigma_\eta = 0.2. \label{EqMeanRevertingBenchmarkFixedParameters}
\end{flalign}
The $s_0$ value is a common configuration in literature using the stationary
standard deviation of the system. When $\mu = \mathcal{L}$ the benchmark sets
the beta parameter as $\sigma_\eta$ to maintain the same stationary variance
between the Gaussian and Laplace cases. The benchmark propagates uncertainty
through the transition model and compares the result against a Monte Carlo
ground truth using the Wasserstein distance. The benchmark evaluates 36
distinct configurations, and it uses the Autocorrelation correlation-tracking
subsystem for all UxHw configurations.

\begin{table}[t]
    \caption[Stochastic volatility transition model benchmark results.]{UxHw configuration
    benchmarks against corresponding equal-accuracy and equal-speed Monte Carlo
    simulation for the \textbf{stochastic volatility transition} model
    (Equation~\ref{EqStochasticVolatilityProcess}). The UxHw configurations in
    this benchmark use the Autocorrelation correlation-tracking subsystem.}
    \label{TableStochasticVolTransitionEMCC}
    \centering
    \footnotesize
    \begin{tabular}{l r r|r r r|r r r}
    \toprule
    \multicolumn{3}{c|}{\textbf{UxHw}}         & \multicolumn{3}{c|}{\textbf{Iso-quality Monte Carlo}} & \multicolumn{3}{c}{\textbf{Iso-time Monte Carlo}} \\
    \midrule
        & Avg\textsuperscript{\textdagger}& Runtime      & Avg Iter. & Runtime       & Speedup Avg         & Att. Iter.              & Avg  & \%$W_1$ change \\
    Method               &     $W_1$      & ($\mu$s)     & Count     & Avg ($\mu$s)  & (MC$\to$UxHw)       & $\lceil$Count$\rceil$   &   $W_1$  & (MC$\to$UxHw) \\
    \midrule
    \rowcolor{a} UxHw 8  & 0.009896       & 814      & 18513.85   & 655094   &  \texttimes 805.18  & 23    & 0.259526 & -96.19\% \\
    \rowcolor{b} UxHw 16 & 0.005762       & 3708     & 55028.43   & 1947126  &  \texttimes 525.16  & 105   & 0.122203 & -95.28\% \\
    \rowcolor{a} UxHw 32 & 0.003702       & 16124    & 133661.02  & 4729462  &  \texttimes 293.32  & 456   & 0.059227 & -93.75\% \\
    \midrule
                         &                &              & \multicolumn{3}{c|}{\textit{(90-th percentile values)}} &\multicolumn{3}{c}{\textit{(90-th percentile values)}} \\
    \cmidrule{4-9}
    \rowcolor{a}\multicolumn{3}{>{\cellcolor{white}}l|}{}& 32273.26  & 1141957   &  \texttimes 1403.58  &    & 0.385836 & -97.44\% \\
    \rowcolor{b}\multicolumn{3}{>{\cellcolor{white}}l|}{}& 95949.39  & 3395073   &  \texttimes 915.68   &    & 0.180736 & -96.81\% \\
    \rowcolor{a}\multicolumn{3}{>{\cellcolor{white}}l|}{}& 232929.04 & 8241961   &  \texttimes 511.16   &    & 0.086084 & -95.70\% \\
    \bottomrule
    \multicolumn{9}{p{0.9\textwidth}}{Avg: average. Iter: iteration. Att:
    attainable. \textsuperscript{\textdagger}UxHw is deterministic and the same
    inputs always yield the same output; the column represents the average
    across benchmark inputs.}
    \end{tabular}
\end{table}

Table~\ref{TableStochasticVolTransitionEMCC} lists the benchmark results. UxHw~8
achieves an average \texttimes 805.18 speedup against the corresponding
iso-quality Monte Carlo simulations, and as much as \texttimes 1403.58 in the
worst-performing 10\% of Monte Carlo batches. UxHw~16 and UxHw~32 achieve
\texttimes 525.16 and \texttimes 293.32 speedup on average, and as much as
\texttimes 915.68 and \texttimes 511.16. The speedup decreases with UxHw
representation size, which is consistent with increased representation and
correlation-tracking overhead.

The comparison against the iso-time Monte Carlo simulations shows smaller
(better) Wasserstein distance for UxHw across all configurations, with a
decreasing relative gap as UxHw runtime increases. UxHw~8 achieves on average
-96.19\% Wasserstein distance in the same latency as the iso-time Monte Carlo
simulations, and as much as -97.44\% in the worst-performing 10\% of Monte Carlo
batches. UxHw~16 and UxHw~32 achieve -95.28\% and -93.75\% Wasserstein distances
on average, and as much as -96.81\% and -95.70\%. This pattern is consistent
with the larger UxHw runtimes at higher representation sizes permitting a larger
attainable Monte Carlo particle count, which reduces the iso-time Monte Carlo
distance and narrows the percentage improvement. The absolute $W_1$ values
decrease with representation size, with average $W_1$ equal to \num{0.009896},
\num{0.005762}, and \num{0.003702} for UxHw~8, UxHw~16, and UxHw~32.

These results show that processor-native uncertainty tracking attains large
speedups and smaller Wasserstein distance for this transition update across the
tested shock families, and that increasing UxHw representation size improves
absolute accuracy while reducing the relative advantage in the iso-time regime
when longer UxHw latency increases the attainable Monte Carlo fidelity.

\subsubsection{Stochastic volatility observation model benchmark}

This benchmark tests the speed and quality of processor-native uncertainty
tracking for the stochastic volatility observation model
(Equation~\ref{EqStochasticVolatilityObservation}) on the UxHw-FPGA-17k hardware
module. Let $X_k$ denote the state uncertainty at step $k$, and let $E_k$ denote
the observation noise term $\varepsilon_k$ in the multiplicative update $Y_k
\coloneqq \exp(X_k/2)\,E_k$. The benchmark uses the configuration
ofEquations~\ref{EqMeanRevertingBenchmarkXkDistribution}--\ref{EqMeanRevertingBenchmarkFixedParameters}
but fixes $X_k$ to a Gaussian family. It tests the usual Gaussian assumption for
$E_k$ and a heavier-tailed alternative~\cite{asai2009bayesian} modelled here as
Laplace noise with matched innovation variance.
Equation~\ref{EqMeanRevertingBenchmarkStateDistribution}--\ref{EqMeanRevertingBenchmarkObservationNoiseScale}
summarise the benchmark configuration space:
\begin{flalign}
    X_k &\sim \mathcal{N}(\lambda_x, s_x), \label{EqMeanRevertingBenchmarkStateDistribution} \\
    \lambda_x &\in \{-2,\,-1,\,-0.5,\,-0.25,\,0,\,0.25,\,0.5,\,1,\,2\}, \text{and} \label{EqMeanRevertingBenchmarkStateMeanValues}\\
    s_x &\in \{1,\, s_0\}, \text{ and} \label{EqMeanRevertingBenchmarkStateVarianceValues}\\
    E_k &\sim \mu(0,\sigma_\varepsilon),\qquad \mu \in \{\mathcal{N},
        \mathcal{L}(b)\}, \text{ with} \label{EqMeanRevertingBenchmarkObservationNoiseDistribution}\\
    b &= \sigma_\varepsilon / \sqrt{2} \text{ when } \mu=\mathcal{L} \text{ so that } \operatorname{Var}(E_k)=\sigma_\varepsilon^2. \label{EqMeanRevertingBenchmarkObservationNoiseScale}
\end{flalign}
The benchmark propagates uncertainty through the observation update and compares
the resulting law against a Monte Carlo ground truth using the Wasserstein
distance. The benchmark evaluates 36 distinct configurations. The UxHw
configurations use the Autocorrelation subsystem.

\begin{table}[t]
    \caption[Stochastic volatility observation model benchmark results.]{UxHw configuration
    benchmarks against corresponding equal-accuracy and equal-speed Monte Carlo
    simulation for the \textbf{stochastic volatility observation} model
    (Equation~\ref{EqStochasticVolatilityObservation}).}
    \label{TableStochasticVolObservationEMCC}
    \centering
    \footnotesize
    \begin{tabular}{l r r|r r r|r r r}
    \toprule
    \multicolumn{3}{c|}{\textbf{UxHw}}         & \multicolumn{3}{c|}{\textbf{Iso-quality Monte Carlo}} & \multicolumn{3}{c}{\textbf{Iso-time Monte Carlo}} \\
    \midrule
        & Avg\textsuperscript{\textdagger}& Runtime      & Avg Iter. & Runtime       & Speedup Avg         & Att. Iter.              & Avg  & \%$W_1$ change \\
    Method               &     $W_1$      & ($\mu$s)     & Count     & Avg ($\mu$s)  & (MC$\to$UxHw)       & $\lceil$Count$\rceil$   &   $W_1$  & (MC$\to$UxHw) \\
    \midrule
    \rowcolor{a} UxHw 8  & 0.074508       & 8985      & 1106.36   & 222156   &  \texttimes 24.73  & 45    & 0.319584 & -76.69\% \\
    \rowcolor{b} UxHw 16 & 0.030924       & 37386     & 6462.86   & 1297737  &  \texttimes 34.71  & 187   & 0.163129 & -81.04\% \\
    \rowcolor{a} UxHw 32 & 0.012192       & 165221    & 41448.50  & 8322826  &  \texttimes 50.37  & 823   & 0.079618 & -84.69\% \\
    \midrule
                         &                &              & \multicolumn{3}{c|}{\textit{(90-th percentile values)}} &\multicolumn{3}{c}{\textit{(90-th percentile values)}} \\
    \cmidrule{4-9}
    \rowcolor{a}\multicolumn{3}{>{\cellcolor{white}}l|}{}& 1745.70   & 350535   &  \texttimes 39.01  &    & 0.597771 & -87.54\% \\
    \rowcolor{b}\multicolumn{3}{>{\cellcolor{white}}l|}{}& 10204.05  & 2048965  &  \texttimes 54.81  &    & 0.302697 & -89.78\% \\
    \rowcolor{a}\multicolumn{3}{>{\cellcolor{white}}l|}{}& 65406.59  & 13133591 &  \texttimes 79.49  &    & 0.149387 & -91.84\% \\
    \bottomrule
    \multicolumn{9}{p{0.9\textwidth}}{Avg: average. Iter: iteration. Att:
    attainable. \textsuperscript{\textdagger}UxHw is deterministic and the same
    inputs always yield the same output; the column represents the average
    across benchmark inputs.}
    \end{tabular}
\end{table}

Table~\ref{TableStochasticVolObservationEMCC} shows a consistent trade-off
between the UxHw tuning parameter $\eta$ and accuracy for the stochastic
volatility observation update. Larger UxHw configurations reduce the average
Wasserstein distance. Against the corresponding iso-quality Monte Carlo
simulations, UxHw achieves average speedups of \texttimes 24.7, \texttimes 34.7,
and \texttimes 50.4, and as much as \texttimes 39.0, \texttimes 54.8, and
\texttimes 79.5 for the worst-performing 10\% MC batches.

The iso-time comparison gives the complementary view, where Monte Carlo uses the
largest attainable iteration budget within the UxHw runtime. Under the
corresponding time budget, UxHw achieves an average $W_1$ distance reduction by
-76.7\%, -81.0\%, and -84.7\%. Larger UxHw configurations yield better
efficiency because the Monte Carlo iteration count required to match UxHw
accuracy grows faster than the UxHw runtime across the same sweep. For the
worst-performing 10\% MC batches, UxHw achieves distance -87.54\%, -89.78\%, and
-91.84\% for UxHw~8, UxHw~16, and UxHw~32, respectively.

These results suggest that the observation update, which combines an exponential
with a multiplicative noise term, produces configurations where Monte Carlo
needs large sample counts to control tail effects, while UxHw maintains
predictable runtime and improves accuracy monotonically with $\eta$.

\subsubsection{Stochastic volatility inference benchmark}

This benchmark tests the speed and accuracy of inference using UxHw BayesLaplace
against the importance sampling algorithm of the particle filter for different
sizes. This benchmark does not use explicit iso-time and iso-quality Monte Carlo
adversaries and instead computes the latency (speed) and accuracy ($W_1$) for
importance sampling inference with specific particle counts. 

\begin{table}[tb]
    \centering
    \caption[Stochastic volatility likelihood model inference benchmark latency
     and accuracy results.]{Stochastic volatility likelihood model inference
     benchmark latency and accuracy results for importance sampling (MC) and
     BayesLaplace (UxHw).}
    \label{TableStochVolInferenceBenchmarkResults}
    \centering
    \footnotesize
    \begin{tabular}{lr|S[table-format=6.0,group-minimum-digits=4]S[table-format=1.6]S[table-format=1.6]}
    \toprule
    Method & Fidelity & {Latency ($\mu$s) \scalebox{0.8}{\rotatebox{90}{\ding{222}}}} & {Mean $W_1$} & {$W_1$ Std} \\
    \midrule
    \rowcolor{a}    MC & 10 & 5793 & 0.482255 & 0.190635 \\
    \rowcolor{b}    MC & 20 & 11651 & 0.351222 & 0.137857 \\
    \rowcolor{a}    UxHw & 8 & 20803 & 0.058566 & 0.019230 \\
    \rowcolor{b}    MC & 50 & 29617 & 0.228117 & 0.089055 \\
    \rowcolor{a}    MC & 100 & 59722 & 0.163482 & 0.062948 \\
    \rowcolor{b}    UxHw & 16 & 72104 & 0.030325 & 0.009997 \\
    \rowcolor{a}    UxHw & 32 & 254353 & 0.022761 & 0.007901 \\
    \rowcolor{b}    MC & 500 & 313324 & 0.075388 & 0.028843 \\
    \rowcolor{a}    MC & 1000 & 631903 & 0.054679 & 0.020856 \\
    \bottomrule
    \end{tabular}
\end{table}

\begin{figure}[tb]
    \centering
    \includegraphics[trim={0cm 0cm 0cm 0.46cm},clip,width=0.95\textwidth]{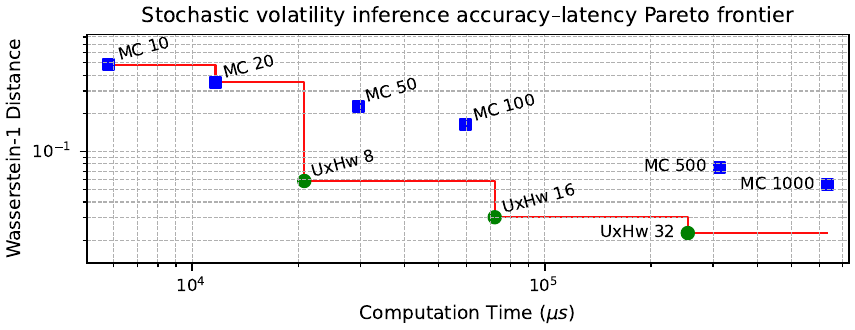}
    \caption[Pareto frontier plot for accuracy and latency of the inference
    mechanism for the Stochastic Volatility system.]{Pareto frontier plot for
    accuracy and latency of the inference mechanism for the Stochastic
    Volatility system. Green circles are UxHw BayesLaplace inference and blue
    squares are inference via importance
    sampling.\label{FigStochVolAccuracyLatencyPareto}}
\end{figure}

Table~\ref{TableStochVolInferenceBenchmarkResults} lists the measured latency and
mean Wasserstein distance error, and Figure~\ref{FigStochVolAccuracyLatencyPareto} shows the
resulting accuracy-latency Pareto frontier. The benchmark compares each inferred
posterior against a Monte Carlo ground truth posterior using the Wasserstein
distance and reports the mean and standard deviation of $W_1$ across all tested
instances.

UxHw~8 attains Wasserstein distance \num{0.058566} within
\qty{20803}{\micro\second}, which dominates importance sampling at 50 and 100
particles by improving both latency and accuracy. UxHw~16 further reduces the
distance to \num{0.030325} within \qty{72104}{\micro\second}, moving the
frontier downward relative to all tested importance sampling sizes. UxHw~32
attains the lowest average $W_1$ distance, \num{0.022761} within
\qty{254353}{\micro\second}, and remains both faster and more accurate than
importance sampling at 500 and 1000 particles. The $W_1$ standard deviation
follows the same trend: importance sampling shows high variability at low
particle counts, whereas UxHw reduces variability as $\eta$ increases.

These results show that UxHw BayesLaplace shifts the stochastic volatility
inference frontier downward, with UxHw~8 covering the mid-latency regime while
substantially reducing distance, and with UxHw~16 and UxHw~32 achieving
high-accuracy posteriors at lower latency than high particle counts.

\subsection{Gordon--Salmond--Smith system}

Let $\upsilon_k$ and $\nu_k$ each denote iid noise.
Equations~\ref{EqGordonWalkTransitionBayesLaplace} and~\ref{EqGordonWalkObservationBayesLaplace} are a
non-linear time-varying system commonly used as a reference system in Bayesian
filtering benchmarking~\cite{gordon1993novel, kitagawa1996monte}. This work
refers to it as the Gordon--Salmond--Smith system. 
\begin{align}
    x_k &= \frac{x_{k-1}}{2} + \frac{25 x_{k-1}}{1 + x_{k-1}^2} + 8 \cos(1.2 k) + \upsilon_k, \label{EqGordonWalkTransitionBayesLaplace}\\
    z_k &= \frac{x_k^2}{20} + \nu_k. \label{EqGordonWalkObservationBayesLaplace}
\end{align}

This section benchmarks the conventional configuration of the
Gordon--Salmond--Smith system of $\upsilon_k \sim \mathcal{N}(0, 3^2)$, $\nu_k
\sim \mathcal{N}(0, 1^2)$, and initial estimate $x_0 \sim \mathcal{N}(0, 5^2)$.
The benchmark uses the UxHw BayesLaplace filter
(Section~\ref{SectionUxHwBayesLaplaceFilter}) for configuration sizes $\eta \in
\{8, 16, 32, 64\}$ against particle filters with counts $n \in \{10, 20, 50,
100, 500, 1000\}$. The benchmark generates 50 traces of the system and each
filter runs on each trace 100 times. It then computes the RMSE across all runs
and traces as a metric of estimation error, the average per-trace RMSE Std as a
metric of estimator stability, and the NLL and the CRPS as posterior quality
metrics.

\begin{table}
\centering
\caption[Gordon--Salmond--Smith system filter accuracy results.]{Estimation
accuracy of different configuration sizes of the UxHw BayesLaplace filter and
the bootstrap particle filter for the \textbf{Gordon--Salmond--Smith system} of
Equations~\ref{EqGordonWalkTransitionBayesLaplace} and~\ref{EqGordonWalkObservationBayesLaplace} with
$\upsilon_k \sim \mathcal{N}(0, 3^2)$, $\nu_k \sim \mathcal{N}(0, 1^2)$,
and initial estimate $x_0 \sim \mathcal{N}(0, 5^2)$.}
\label{TableGordonFilterAccuracy}
\footnotesize
\begin{tabular}{lr|S[table-format=1.6]S[table-format=1.6]S[table-format=3.6]S[table-format=1.6]}
\toprule
\textbf{Method} & \textbf{\#Traces} & \textbf{RMSE}\enspace\scalebox{0.8}{\rotatebox{90}{\ding{222}}} & \textbf{Per-trace RMSE Std} & \textbf{NLL} & \textbf{CRPS} \\
\midrule
\rowcolor{a} UxHw 64  &       50 & 4.554662 &           0.000000 &   2.385085 & 1.722927 \\
\rowcolor{b} BPF 1000 &       50 & 4.631633 &           0.121110 &   2.743835 & 1.753481 \\
\rowcolor{a} BPF 500  &       50 & 4.649263 &           0.166242 &   3.125517 & 1.765093 \\
\rowcolor{b} UxHw 32  &       50 & 4.657051 &           0.000000 &   2.811172 & 1.775729 \\
\rowcolor{a} UxHw 16  &       50 & 4.994060 &           0.000000 &   4.308073 & 1.968334 \\
\rowcolor{b} BPF 100  &       50 & 5.016186 &           0.635257 &   9.264011 & 1.964372 \\
\rowcolor{a} BPF 50   &       50 & 5.643520 &           1.073605 &  23.979015 & 2.330302 \\
\rowcolor{b} UxHw 8   &       50 & 6.271464 &           0.000000 &   8.375627 & 2.754201 \\
\rowcolor{a} BPF 20   &       50 & 7.202361 &           1.487972 &  77.521031 & 3.467684 \\
\rowcolor{b} BPF 10   &       50 & 8.617990 &           1.539336 & 150.995513 & 4.790006 \\
\bottomrule
\end{tabular}
\end{table}

\begin{figure}[tb]
    \centering
    \includegraphics[trim={0.1cm 0.1cm 0.1cm 0cm},clip,width=0.9\textwidth]{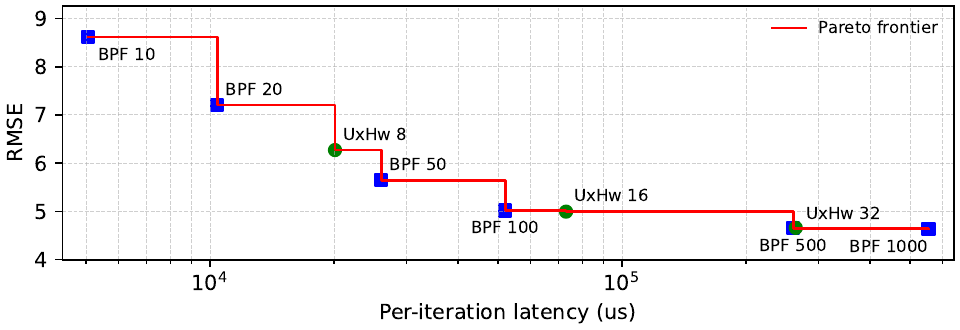}
    \caption[Pareto frontier plot for filtering RMSE and latency of the
    Gordon--Salmond--Smith system.]{Pareto frontier plot for filtering RMSE and
    latency of the Gordon--Salmond--Smith system. Green circles are the
    BayesLaplace filter and blue squares are the bootstrap particle filter. The
    UxHw-based filter achieves head-to-head performance with the particle filter
    in raw point-estimate error.\label{FigGordonSalmondSmithRMSELatencyPareto}}
\end{figure}

Table~\ref{TableGordonFilterAccuracy} presents the filtering benchmark results
sorted by ascending RMSE. The lowest estimation error comes from UxHw~64, which
achieves a cross-trace cross-run average RMSE of \num{4.554662}. The
1000-particle bootstrap filter follows with RMSE \num{4.631633}, and the
500-particle variant reaches \num{4.649263}. UxHw~32 achieves RMSE
\num{4.657051}, which places it close to the largest particle filters. UxHw~16
and UxHw~8 degrade more clearly, with RMSE \num{4.994060} and \num{6.271464},
respectively. As expected, the particle filter improves monotonically with
particle count, and the returns diminish between 500 and 1000 particles. The
UxHw filters also improve monotonically with configuration size in this
benchmark, with UxHw~64 giving the strongest point-estimate accuracy.
Figure~\ref{FigGordonSalmondSmithRMSELatencyPareto} shows the same head-to-head
behaviour on the RMSE--latency frontier, where the strongest UxHw and
particle-filter configurations lie close in raw point-estimate error.

The per-trace RMSE standard deviation indicates estimation stability. The UxHw
filters rely on deterministic distributional arithmetic, so repeated runs on the
same trace yield identical estimates up to numerical precision in the order of
\num{e-17}. In contrast, the particle filters exhibit run-to-run variability
from stochastic propagation and resampling. This variability increases as the
particle count decreases, with per-trace RMSE standard deviation rising from
\num{0.121110} at 1000 particles to \num{1.539336} at 10 particles. This
contrast makes the UxHw filters attractive when repeatability is an important
requirement.

The negative log-likelihood score measures the probability density that the
posterior assigns to the true latent state, and lower is better. UxHw~64
achieves the best NLL of \num{2.385085}. The next best results come from
BPF~1000 at \num{2.743835}, UxHw~32 at \num{2.811172}, and BPF~500 at
\num{3.125517}. Smaller filters degrade more clearly. UxHw~16 reaches
\num{4.308073}, and UxHw~8 reaches \num{8.375627}. In contrast, the
particle-filter variants below 100 particles deteriorate strongly, with NLL
\num{9.264011} at 100 particles, \num{23.979015} at 50 particles,
\num{77.521031} at 20 particles, and \num{150.995513} at 10 particles. These
results show that the largest UxHw configuration gives the sharpest posterior at
the true state in this benchmark and that the smallest UxHw configuration
retains strong information compared to similar-latency particle filter
configurations.

The continuous ranked probability score measures calibration and sharpness
around the true latent state, penalising over-spread posteriors, and lower is
better. UxHw~64 again achieves the best result, with CRPS \num{1.722927}. The
1000-particle and 500-particle bootstrap filters follow with \num{1.753481} and
\num{1.765093}, and UxHw~32 reaches \num{1.775729}. BPF~100 and UxHw~16 are very
close, with CRPS \num{1.964372} and \num{1.968334}, respectively. The
lower-fidelity methods degrade more clearly, reaching \num{2.330302} for BPF~50,
\num{2.754201} for UxHw~8, \num{3.467684} for BPF~20, and \num{4.790006} for
BPF~10. The ranking by CRPS broadly tracks the ranking by RMSE for the strongest
methods. Overall, UxHw~64 and BPF 1000 are the strongest methods in this
benchmark, with UxHw~64 giving the best accuracy while also preserving
deterministic repeatability.

The following subsections dive deeper into the constituent parts of the
BayesLaplace filter: the transition model, the observation model, and the
inference mechanism, and make side-by-side benchmarking with the corresponding
parts of the particle filter algorithm.

\subsubsection{Gordon--Salmond--Smith transition model benchmark}

This benchmark tests the speed and quality of processor-native uncertainty
tracking for the Gordon--Salmond--Smith transition equation using the
UxHw-FPGA-17k hardware module. The benchmark tests the Wasserstein distance of
the uncertainty propagation result for nine hand-picked high-likelihood state
locations. Let $X_{k-1}$ denote the random variable representing the previous
state. For each state location value,
Equations~\ref{EqGSSStateDistribution}--\ref{EqGSSPerturbationTerm} summarise
the benchmark configuration space:
\begin{flalign}
    X_{k-1} &\sim \mu(\lambda_x, 1), \text{with}\label{EqGSSStateDistribution}\\
    \lambda_x &\in \{-15,\,-12,\,-1,\,-0.5,\,0,\,0.5,\,1,\,12,\,15\}, \text{and}\\
    \mu &\in \{\mathcal{N}, \mathcal{U}, \mathcal{L} \}, \text{and}\\
    \Upsilon_k &\sim \mathcal{N}(0, 3). \label{EqGSSPerturbationTerm}
\end{flalign}
The benchmark evaluates three parametric distribution families for the state
uncertainty and adds an independent Gaussian perturbation term consistent with
the reference model. The benchmark then propagates uncertainty through the
transition equation under test using $(X_{k-1}, \Upsilon_k)$ as inputs, and
compares the resulting law against an Monte Carlo ground truth using the
Wasserstein distance. Overall, the benchmark evaluates 27 distinct
configurations.

\begin{table}[t]
    \caption[Gordon--Salmond--Smith transition
    model benchmark results.]{UxHw configuration benchmarks against corresponding equal-accuracy
    and equal-speed Monte Carlo simulation for the
    \textbf{Gordon--Salmond--Smith transition} model
    (Equation~\ref{EqGordonWalkTransitionBayesLaplace}). The UxHw configurations in this
    benchmark use the Autocorrelation correlation-tracking subsystem.}
    \label{TableGSSTransitionEMCC}
    \centering
    \footnotesize
    \begin{tabular}{l r r|r r r|r r r}
    \toprule
    \multicolumn{3}{c|}{\textbf{UxHw}}         & \multicolumn{3}{c|}{\textbf{Iso-quality Monte Carlo}} & \multicolumn{3}{c}{\textbf{Iso-time Monte Carlo}} \\
    \midrule
        & Avg\textsuperscript{\textdagger}& Runtime      & Avg Iter. & Runtime       & Speedup Avg         & Att. Iter.              & Avg  & \%$W_1$ change \\
    Method               &     $W_1$      & ($\mu$s)     & Count     & Avg ($\mu$s)  & (MC$\to$UxHw)       & $\lceil$Count$\rceil$   &   $W_1$  & (MC$\to$UxHw) \\
    \midrule
    \rowcolor{a} UxHw 8  & 0.133420       & 991        & 5521.90    & 414776   &  \texttimes 418.59  & 14      & 1.749511 & -92.37\% \\
    \rowcolor{b} UxHw 16 & 0.046058       & 3719       & 34942.25   & 2624680  &  \texttimes 705.83  & 50      & 0.936065 & -95.08\% \\
    \rowcolor{a} UxHw 32 & 0.017634       & 16226      & 169966.90  & 12767030 &  \texttimes 786.82  & 217     & 0.451648 & -96.10\% \\
    \midrule
                         &                &              & \multicolumn{3}{c|}{\textit{(90-th percentile values)}} &\multicolumn{3}{c}{\textit{(90-th percentile values)}} \\
    \cmidrule{4-9}
    \rowcolor{a}\multicolumn{3}{>{\cellcolor{white}}l|}{}& 9778.33 & 734497     &   \texttimes 741.25   &    & 3.308902 & -95.97\% \\
    \rowcolor{b}\multicolumn{3}{>{\cellcolor{white}}l|}{}& 62433.98 & 4689716   &   \texttimes 1261.16  &    & 1.804655 & -97.45\% \\
    \rowcolor{a}\multicolumn{3}{>{\cellcolor{white}}l|}{}& 307733.59 & 23115347   &  \texttimes 1424.58 &    & 0.862063 & -97.95\% \\
    \bottomrule
    \multicolumn{9}{p{0.9\textwidth}}{Avg: average. Iter: iteration. Att:
    attainable. \textsuperscript{\textdagger}UxHw is deterministic and the same
    inputs always yield the same output; the column represents the average
    across benchmark inputs.}
    \end{tabular}
\end{table}

\begin{figure}[tbp]
\centering
\includegraphics[trim={0.1cm 0.1cm 0.1cm 0.9cm},clip,width=\textwidth]{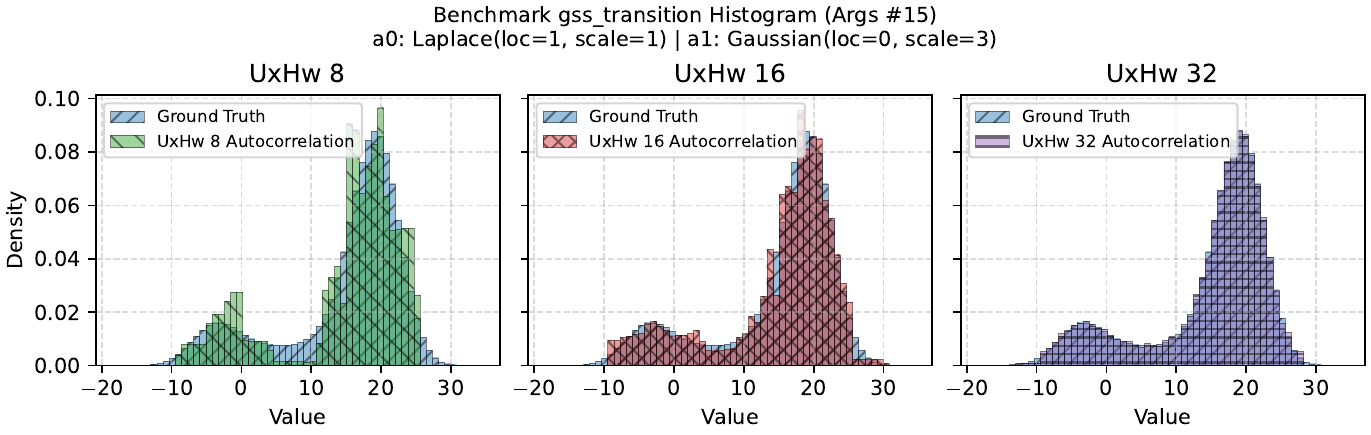}
\caption[Comparison of uncertainty-tracking output for Gordon--Salmond--Smith
transition for different UxHw representation sizes.]{Comparison of
uncertainty-tracking output for Gordon--Salmond--Smith transition of
Equation~\ref{EqGordonWalkTransitionBayesLaplace} using the UxHw with $\eta \in
\{8, 16, 32\}$ and with the Autocorrelation subsystem, for $X_{k-1} \sim
\mathcal{L}(1, 1)$ and $\Upsilon_k \sim \mathcal{N}(0, 3^2)$.
\label{FigInferenceGSSTransition}}
\end{figure}

Table~\ref{TableGSSTransitionEMCC} shows the results for the
Gordon--Salmond--Smith transition model benchmark.
Figure~\ref{FigInferenceGSSTransition} shows the uncertainty-tracking output of
the transition model for $X_{k-1} \sim \mathcal{L}(1, 1)$ and $\Upsilon_k \sim
\mathcal{N}(0, 3^2)$. UxHw~8 achieves an average \texttimes 418.59 speedup
against the corresponding iso-quality Monte Carlo simulations, and as much as
\texttimes 741.25 in 10\% of the benchmark cases. UxHw~16 and UxHw~32 achieve
\texttimes 705.83 and \texttimes 786.82 speedup on average, and as much as
\texttimes 1261.16 and \texttimes 1424.58 in 10\% of the benchmark cases where
Monte Carlo performs worst. The speedup grows with representation size because
the benchmark transition equation increases both nonlinearity and correlation
between intermediate quantities, and the Autocorrelation subsystem keeps this
dependence information within the UxHw execution.

The comparison against the iso-time Monte Carlo simulations follows the same
ordering across UxHw configurations. UxHw~8 achieves on average -92.37\%
Wasserstein distance in the same latency as the iso-time Monte Carlo
simulations, and as much as -95.97\% for 10\% of the benchmark cases. UxHw~16
and UxHw~32 achieve -95.08\% and -96.10\% Wasserstein distances on average, and
as much as -97.45\% and -97.95\% for the worst-performing 10\% of the Monte
Carlo cases. The absolute $W_1$ values decrease with representation size, with
average $W_1$ equal to \num{0.133420}, \num{0.046058}, and \num{0.017634} for
UxHw~8, UxHw~16, and UxHw~32, which matches the additional resolution available
for capturing the multimodality and tail mass induced by the transition
equation.

The result show the highest improvement coefficient between UxHw~8 and UxHw~16
which signals that this change in UxHw representation size allowed the system to
better capture the nonlinearity of the model.

\begin{table}[t]
    \caption[Gordon--Salmond--Smith observation
    model benchmark results.]{UxHw configuration benchmarks against corresponding equal-accuracy
    and equal-speed Monte Carlo simulation for the
    \textbf{Gordon--Salmond--Smith observation} model
    (Equation~\ref{EqGordonWalkObservationBayesLaplace}). The UxHw configurations in this
    benchmark use the Autocorrelation correlation-tracking subsystem.}
    \label{TableGSSlObservationEMCC}
    \centering
    \footnotesize
    \begin{tabular}{l r r|r r r|r r r}
    \toprule
    \multicolumn{3}{c|}{\textbf{UxHw}}         & \multicolumn{3}{c|}{\textbf{Iso-quality Monte Carlo}} & \multicolumn{3}{c}{\textbf{Iso-time Monte Carlo}} \\
    \midrule
        & Avg\textsuperscript{\textdagger}& Runtime      & Avg Iter. & Runtime       & Speedup Avg         & Att. Iter.              & Avg  & \%$W_1$ change \\
    Method               &     $W_1$      & ($\mu$s)     & Count     & Avg ($\mu$s)  & (MC$\to$UxHw)       & $\lceil$Count$\rceil$   &   $W_1$  & (MC$\to$UxHw) \\
    \midrule
    \rowcolor{a} UxHw 8  & 0.116502       & 317       & 4707.53   & 190378   &  \texttimes 601.11  & 8     & 1.118450 & -89.58\% \\
    \rowcolor{b} UxHw 16 & 0.056425       & 1061      & 12968.77  & 524473   &  \texttimes 494.34  & 27    & 0.638450 & -91.16\% \\
    \rowcolor{a} UxHw 32 & 0.026762       & 4223      & 50872.34  & 2057338  &  \texttimes 487.17  & 105   & 0.330842 & -91.91\% \\
    \midrule
                         &                &              & \multicolumn{3}{c|}{\textit{(90-th percentile values)}} &\multicolumn{3}{c}{\textit{(90-th percentile values)}} \\
    \cmidrule{4-9}
    \rowcolor{a}\multicolumn{3}{>{\cellcolor{white}}l|}{}& 8192.65 & 331321     &  \texttimes 1046.14  &    & 2.444436 & -95.23\% \\
    \rowcolor{b}\multicolumn{3}{>{\cellcolor{white}}l|}{}& 22648.12 & 915917   &  \texttimes  863.30   &    & 1.402285 & -95.98\% \\
    \rowcolor{a}\multicolumn{3}{>{\cellcolor{white}}l|}{}& 88604.08 & 3583255   &  \texttimes 848.50   &    & 0.725952 & -96.31\% \\
    \bottomrule
    \multicolumn{9}{p{0.9\textwidth}}{Avg: average. Iter: iteration. Att:
    attainable. \textsuperscript{\textdagger}UxHw is deterministic and the same
    inputs always yield the same output; the column represents the average
    across benchmark inputs.}
    \end{tabular}
\end{table}

\subsubsection{Gordon--Salmon--Smith observation model benchmark}

This benchmark tests the speed and quality of UxHw-based uncertainty tracking
for the Gordon--Salmond--Smith observation equation using the UxHw-FPGA-17k
hardware module. Let $X_k$ denote the state uncertainty input to the observation
model. The benchmark evaluates the same nine state locations from the
specification in
Equations~\ref{EqGSSStateDistribution}--\ref{EqGSSPerturbationTerm}. Let $\mu$
denote a distribution family in $\{\mathcal{N},\,\mathcal{U},\,\mathcal{L}\}$,
and let $s_x$ denote the scale parameter. The benchmark sets
\begin{flalign}
    X_k &\sim \mu(\lambda_x, s_x), \quad s_x = 3, \text{ and}\nonumber\\
    \nu_k &\sim \mathcal{N}(0, 1), \nonumber
\end{flalign}
and propagates uncertainty through the observation equation. The benchmark
compares the resulting law against a Monte Carlo ground truth using the
Wasserstein distance, and it evaluates 27 distinct configurations.

Table~\ref{TableGSSlObservationEMCC} reports the benchmark results. UxHw~8
achieves an average \texttimes 601.11 speedup against the corresponding
iso-quality Monte Carlo simulations, and as much as \texttimes 1046.14 in 10\%
of the benchmark cases. UxHw~16 and UxHw~32 achieve \texttimes 494.34 and
\texttimes 487.17 speedup on average, and as much as \texttimes 863.30 and
\texttimes 848.50 in 10\% of the benchmark cases where Monte Carlo performs
worst. The average speedup decreases with representation size because the
observation equation has low arithmetic depth while the runtime increases with
UxHw representation size.

The comparison against the iso-time Monte Carlo simulations shows consistently
better output quality for UxHw across configurations. UxHw~8 achieves on average
-89.58\% Wasserstein distance in the same latency as the iso-time Monte Carlo
simulations, and as much as -95.23\% in 10\% of the MC batches. UxHw~16 and
UxHw~32 achieve -91.16\% and -91.91\% Wasserstein distances on average, and as
much as -95.98\% and -96.31\% for the worst-performing 10\% of the Monte Carlo
cases. The absolute $W_1$ values decrease with representation size, with average
$W_1$ equal to \num{0.116502}, \num{0.056425}, and \num{0.026762} for UxHw~8,
UxHw~16, and UxHw~32, respectively.

These results show that processor-native uncertainty tracking achieves large
speedups and smaller Wasserstein distance for the observation equation across
all tested input families, and the quality gain increases modestly with UxHw
representation size while the speedup saturates when the kernel cost becomes
small relative to representation and correlation-tracking overhead.

\subsubsection{Gordon--Salmon--Smith inference benchmark}

This benchmark tests the speed and accuracy of inference using UxHw BayesLaplace
against the importance sampling algorithm from the particle filter for different
sizes. This benchmark does not use explicit iso-time and iso-quality Monte Carlo
adversaries and instead computes the latency (speed) and accuracy ($W_1$) for
importance sampling inference with specific particle counts. 

\begin{table}[tb]
    \centering
    \caption[Gordon--Salmond--Smith likelihood model inference benchmark latency
    and accuracy results.]{Gordon--Salmon--Smith likelihood model inference benchmark latency and
    accuracy results for importance sampling (MC) and BayesLaplace (UxHw).}
    \label{TableGSSInferenceBenchmarkResults}
    \centering
    \footnotesize
    \begin{tabular}{lr|S[table-format=6.0,group-minimum-digits=4]S[table-format=1.6]S[table-format=1.6]}
    \toprule
    Method & Fidelity & {Latency ($\mu$s) \scalebox{0.8}{\rotatebox{90}{\ding{222}}}} & {Mean $W_1$} & {$W_1$ Std} \\
    \midrule
    \rowcolor{a}    MC & 10   & 4298 & 0.918780  & 0.600507 \\
    \rowcolor{b}    MC & 20   & 8915 & 0.655264   & 0.425093 \\
    \rowcolor{a}    UxHw & 8  & 19110 & 0.226054  & 0.101888 \\
    \rowcolor{b}    MC & 50   & 22286 & 0.419784  & 0.270830 \\
    \rowcolor{a}    MC & 100  & 44573 & 0.301244  & 0.191753 \\
    \rowcolor{b}    UxHw & 16 & 69406 & 0.096282  & 0.048861 \\
    \rowcolor{a}    MC & 500  & 222863 & 0.140070 & 0.087858 \\
    \rowcolor{b}    UxHw & 32 & 247796 & 0.052341 & 0.037639 \\
    \rowcolor{a}    MC & 1000 & 479288 & 0.101370 & 0.063020 \\
    \bottomrule
    \end{tabular}
\end{table}

\begin{figure}[tb]
    \centering
    \includegraphics[trim={0cm 0cm 0cm 0.46cm},clip,width=0.95\textwidth]{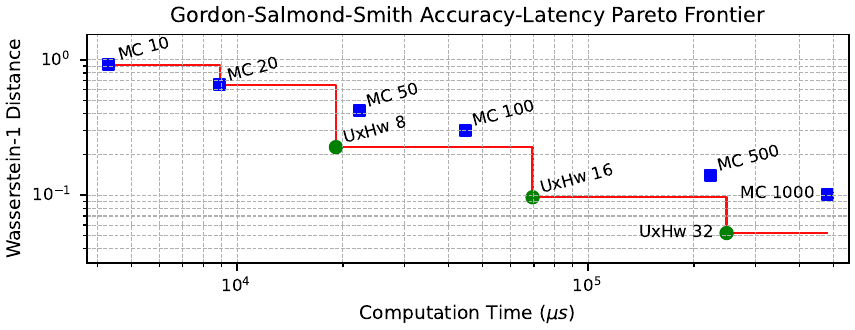}
    \caption[Pareto frontier plot for accuracy and latency of the inference
    mechanism for the Gordon--Salmond--Smith system.]{Pareto frontier plot for
    accuracy and latency of the inference mechanism for the
    Gordon--Salmond--Smith system. Green circles are UxHw BayesLaplace inference
    and blue squares are inference via importance
    sampling.\label{FigGSSAccuracyLatencyPareto}}
\end{figure}

The benchmark constructs inference inputs from a trace of the
Gordon--Salmond--Smith system with 50 times steps and static random seed for
reproducibility. Let $\{x_k\}_{k=1}^{50}$ denote the resulting state sequence
obtained by iterating the ideal transition equation and adding an independent
Gaussian perturbation term with scale $s_x$. The benchmark rounds each $x_k$ to
one decimal place to use it as a prior location parameter. For each such prior
location, the benchmark draws five observations using Gordon--Salmond--Smith
observation model, retaining observations whose likelihood under the
corresponding prior exceeds a fixed threshold 0.1. This keeps the benchmark
focused on high-likelihood inference cases. Let
$\{\{z_{k,i}\}_{i=1}^5\}_{k=1}^{50}$ denote observation values. The benchmark
defines the input as a prior law $X_k$ as either Gaussian or Laplace with
location $x_k$ and scale $s_x$, with the observations $\{z_{k,i}\}_{i=1}^5$.
This benchmark configuration yields 500 distinct inference instances across the
two prior families.

For each inference instance, the benchmark computes the posterior law using the
BayesLaplace operation and using importance sampling with fidelity of $n \in
[10, 20, 50, 100, 500, 1000]$ (particles).
Table~\ref{TableGSSInferenceBenchmarkResults} lists the results for the tested
configurations, and Figure~\ref{FigGSSAccuracyLatencyPareto} shows the resulting
accuracy-latency Pareto frontier. The benchmark compares each inferred posterior
against a Monte Carlo ground truth posterior using the Wasserstein distance
and reports the mean and standard deviation of $W_1$ across all instances
alongside the measured latency. UxHw~8 attains \num{0.226054} with
\qty{19110}{\micro\second} latency, which improves both accuracy and latency
relative to importance sampling with 50 and 100 particles. UxHw~16 attains
\num{0.096282} within \qty{69406}{\micro\second}, which dominates the
high-fidelity importance sampling cases, with lower distance than 1000 particles
at much lower latency. UxHw~32 further reduces the average distance to
\num{0.052341} within \qty{247796}{\micro\second}, which improves accuracy
relative to all tested importance sampling fidelities while remaining faster
than 1000 particles. The standard deviation of $W_1$ follows the same trend,
with importance sampling exhibiting larger variability at low $n$ and UxHw
reducing variability as $\eta$ increases.

These results show that BayesLaplace pushes the inference accuracy-latency
frontier downward for this benchmark, with UxHw~8 matching the latency regime of
mid-fidelity importance sampling while achieving lower $W_1$, and with UxHw~16
and UxHw~32 attaining accuracy comparable to, or better than, high particle
counts but with better latency. The UxHw BayesLaplace inference approach
dominates the importance sampling points and the trend suggests that
BayesLaplace is more efficient at approximating the posterior distribution.

\subsection{Mean-reverting process with exponential heteroscedastic observation}

Mean-reverting processes are common in modelling variables that fluctuate around
an equilibrium on short to medium timescales. This includes homeostatic
physiological variables, thermal dynamics around steady operating points in
electronics, and communications link-quality
metrics~\cite{aalen2004survival,chin2014wireless}. This section uses an
abstracted formulation of a mean-reverting process with cubic feedback as a
benchmark for the UxHw BayesLaplace filter.

Let $x_k$ be the current at time index $k$ and let $c$ be the system-equilibrium
target. Let $\alpha$ and $\beta$ be parameters controlling the mean-reverting
feedback effect of the model, for the linear and the cubic component,
respectively. And, let $\upsilon_k \sim \mu_\upsilon$ denote additive noise. Let
$y_k$ denote the observation and let $m_k \sim \mu_m$ be a multiplicative noise
source. Let $h_k \sim \mu_h(x_k, m_k)$ be heteroscedastic additive noise whose
scale $\sigma_h$ is dependent on the signal value, with base scale
$\sigma_{\mathrm{base}}$. Equations~\ref{EqCubicRevertingProcess}
and~\ref{EqBeerLambertObservationModel} give the state-space equations for this
system:
\begin{align}
    x_{k} &= x_{k-1} + \alpha(c - x_{k-1}) - \alpha\beta(c - x_{k-1})^3 + \upsilon_k, \label{EqCubicRevertingProcess}\\
    y_k &= e^{-x_k}(1 + m_k) + h_k,\quad \text{where} \label{EqBeerLambertObservationModel}\\
    \sigma_{h_k} &\coloneqq \sqrt{\sigma_{\mathrm{base}}^2 + e^{-x_k}(1 + m_k)}\nonumber
\end{align}
The scale of the uncertain term $h_k$ is bigger when the rest of the observation
value is bigger.

This benchmark uses this system to compare UxHw BayesLaplace filter against
conventional bootstrap particle filtering. The following sections examine the
uncertainty-tracking performance of UxHw for the transition model, the
observation model, and the inference step of this system.

This filtering benchmark initially tests this mean-reverting--exponential system
with $\upsilon_k \sim \mathcal{N}(0, 0.05^2)$, $m_k \sim \mathcal{N}(0,
0.05^2)$, $h_k \sim \mathcal{L}(0, \sigma_{h_k})$, $\sigma_{\mathrm{base}} =
0.01$ $c = 0.95$, $\alpha=0.1$, and $\beta=0.05$ using the UxHw BayesLaplace
filter of Section~\ref{SectionUxHwBayesLaplaceFilter} for configuration sizes
$\eta \in \{8, 16, 32, 64\}$ against particle filters with particle counts $n
\in \{10, 20, 50, 100, 500, 1000\}$. The benchmark generates 50 traces of the
system and each filter runs on each trace 100 times. The benchmark computes the
average RMSE across all runs and traces as a metric of estimation error, the
average per-trace RMSE standard deviation as a metric of estimator stability,
and NLL and the CRPS as posterior quality metrics.

\begin{table}
\centering
\caption[Cubic-mean-reverting--exponential-heteroscedastic system filter
accuracy results.]{Estimation accuracy of different configuration sizes of the
UxHw BayesLaplace filter and the bootstrap particle filter for the
\textbf{mean-reverting--exponential} system of
Equations~\ref{EqCubicRevertingProcess} and~\ref{EqBeerLambertObservationModel}
with $\upsilon_k \sim \mathcal{N}(0, 0.05^2)$, $m_k \sim \mathcal{N}(0,
0.05^2)$, $h_k \sim \mathcal{L}(0, \sigma_{h_k})$, $\sigma_{\mathrm{base}} =
0.01$, and initial estimate $x_0 \sim \mathcal{N}(0, 0.1^2)$.}
\label{TableCubicBeerFilterAccuracy}
\footnotesize
\begin{tabular}{lr|S[table-format=1.6]S[table-format=1.6]S[table-format=-1.6]S[table-format=1.6]}
\toprule
\textbf{Method} & \textbf{\#Traces} & \textbf{RMSE}\enspace\scalebox{0.8}{\rotatebox{90}{\ding{222}}} & \textbf{Per-trace RMSE Std} & \textbf{NLL} & \textbf{CRPS} \\
\midrule
\rowcolor{a} BPF 1000 &       50 & 0.109398 &           0.002258 & -0.772228 & 0.062809 \\
\rowcolor{b} UxHw 64  &       50 & 0.109505 &           0.000000 & -0.723642 & 0.062816 \\
\rowcolor{a} BPF 500  &       50 & 0.109620 &           0.003112 & -0.766055 & 0.062968 \\
\rowcolor{b} BPF 100  &       50 & 0.111778 &           0.006678 & -0.714208 & 0.064579 \\
\rowcolor{a} UxHw 32  &       50 & 0.111926 &           0.000000 & -0.547223 & 0.064762 \\
\rowcolor{b} UxHw 16  &       50 & 0.114206 &           0.000000 & -0.404290 & 0.066548 \\
\rowcolor{a} BPF 50   &       50 & 0.114322 &           0.009559 & -0.643363 & 0.066502 \\
\rowcolor{b} UxHw 8   &       50 & 0.114852 &           0.000000 & -0.595478 & 0.067312 \\
\rowcolor{a} BPF 20   &       50 & 0.121551 &           0.014104 & -0.291816 & 0.072266 \\
\rowcolor{b} BPF 10   &       50 & 0.129010 &           0.017815 &  0.781573 & 0.079262 \\
\bottomrule
\end{tabular}
\end{table}

\begin{figure}[tb]
    \centering
    \includegraphics[trim={0.1cm 0.1cm 0.1cm 0cm},clip,width=0.9\textwidth]{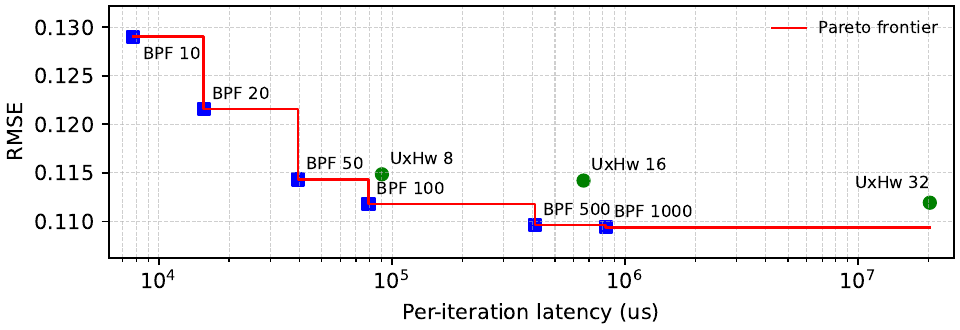}
    \caption[Pareto frontier plot for filtering RMSE and latency of the
    mean-reverting--exponential system.]{Pareto frontier plot for filtering RMSE
    and latency of the cubic mean-reverting transition and exponential
    observation system. Green circles are the BayesLaplace filter and blue
    squares are the bootstrap particle filter. The small latency of UxHw~8
    allows it to stay competitive with the particle
    filter.\label{FigBeerLambertRMSELatencyPareto}}
\end{figure}

Table~\ref{TableCubicBeerFilterAccuracy} presents the filtering benchmark
results sorted by RMSE. The lowest estimation error comes from the 1000-particle
bootstrap filter, which achieves a cross-trace cross-run average RMSE of
\num{0.109398}. UxHw~64 follows close with RMSE \num{0.109505}, and the
500-particle bootstrap filter reaches \num{0.109620}. The next pair is BPF~100
at \num{0.111778} and UxHw~32 at \num{0.111926}. UxHw~16 reaches \num{0.114206},
slightly ahead of BPF~50 at \num{0.114322}, while UxHw~8 reaches \num{0.114852}.
As expected for the particle filter, RMSE improves monotonically with particle
count. The UxHw filters also improve with configuration size in this benchmark,
with UxHw~64 giving the strongest result among the UxHw configurations.

Figure~\ref{FigBeerLambertRMSELatencyPareto} shows the Pareto frontier plot for
the RMSE and per-iteration latency of the benchmarked filters. For this system,
in raw point-estimate accuracy via RMSE, the particle filter maintains
accuracy--latency coefficient better than the UxHw-based approach. The small
latency of UxHw~8 enables it to stay competitive the region around RMSE 0.115
and per-iteration latency \qty{17}{\milli\second}.

The per-trace RMSE standard deviation indicates estimation stability. The UxHw
filters rely on deterministic distributional arithmetic, so repeated runs on the
same trace yield identical estimates up to numerical precision. The results are
in the order of \num{1e-16} to \num{1e-17}. The particle filter shows visible
run-to-run variation, with per-trace RMSE standard deviation increasing as the
particle count decreases. This contrast makes the UxHw filters attractive when
repeatability is an important requirement.

In terms of negative log-likelihood, the 1000-particle bootstrap filter achieves
the best NLL of \num{-0.772228}, followed by the 500-particle filter at
\num{-0.766055}. UxHw~64 is the strongest UxHw configuration with NLL
\num{-0.723642}, and BPF~100 follows closely at \num{-0.714208}. Among the
remaining methods, BPF~50 reaches \num{-0.643363}, UxHw~8 reaches
\num{-0.595478}, UxHw~32 reaches \num{-0.547223}, and UxHw~16 reaches
\num{-0.404290}. These results show that the particle filter gives sharper
posterior density at the true state in this benchmark, while UxHw~64 remains
competitive with the stronger particle-filter configurations.

The continuous ranked probability score measures calibration and sharpness
around the true latent state, penalising over-spread posteriors, and lower is
better. The 1000-particle bootstrap filter achieves the best CRPS of
\num{0.062809}, with UxHw~64 essentially tied at \num{0.062816}. The next best
results come from BPF~500 at \num{0.062968}, BPF~100 at \num{0.064579}, and
UxHw~32 at \num{0.064762}. UxHw~16 and BPF~50 remain close at \num{0.066548}
and \num{0.066502}, respectively, while UxHw~8 reaches \num{0.067312}. The
ranking by CRPS broadly follows the ranking by RMSE for the strongest methods.
Overall, the results identify BPF~1000 and UxHw~64 as the strongest
configurations in this benchmark, with UxHw~64 remaining close to the best
particle-filter result while preserving deterministic repeatability.

The following subsections dive deeper into the constituent parts of the
UxHw-based BayesLaplace filter: the transition model, the observation model, and
the inference mechanism, and make side-by-side benchmarking with the
corresponding parts of the particle filter algorithm.

\subsubsection{Cubic mean-reverting transition model benchmark}
\label{SectionCubicRevertingBenchmark}

Let $X_k$ be the random variable corresponding to $x_k$, and $\Upsilon_k$ to
$\upsilon_k$. The benchmark tests the speed and quality of UxHw uncertainty
tracking for the cubic mean-reverting transition model of
Equation~\ref{EqCubicRevertingProcessTransformed} on the UxHw-FPGA-17k hardware
module. This is Equation~\ref{EqCubicRevertingProcess} adapted for better
uncertainty tracking on the UxHw system by rewriting the double subtraction
from $c$.
\begin{equation}
    X_{k} = X_{k-1} + \alpha \left(c - X_{k-1}\right) \left(1.0 + \beta (c - X_{k-1})^2\right) + \Upsilon_k. \label{EqCubicRevertingProcessTransformed}
\end{equation}

Equations~\ref{EqBenchmarkStateInit}--\ref{EqBenchmarkNoiseVar} summarise the
benchmark configuration space:
\begin{flalign}
    X_{k-1} &\sim \mathcal{N}(\lambda_x, s_x), \label{EqBenchmarkStateInit}\\
    \lambda_x &\in \{0.55,\,0.8,\,0.9,\,0.95,\,1.0,\,1.1,\,1.45\}, \text{ and}\\
    s_x &\in \{0.1,\,0.15\}, \text{ and}\\
    \Upsilon_k &\sim \mu_\Upsilon(0,\sigma_\upsilon),\qquad \mu_\Upsilon \in
    \{\mathcal{N},\,\mathcal{L}(b),\,\mathcal{U}(w)\}, \text{ with}\\
    \sigma_\upsilon &\in \{0.05,\,0.10\}, \text{ and}\\
    b &= \sigma_\upsilon / \sqrt{2} \text{ when } \mu_\Upsilon=\mathcal{L} \text{ so that }
    \operatorname{Var}(\Upsilon_k)=\sigma_\upsilon^2, \text{ and}\\
    w &= 2\sqrt{3}\,\sigma_\upsilon \text{ when } \mu_\Upsilon=\mathcal{U} \text{ so that }
    \operatorname{Var}(\Upsilon_k)=\sigma_\upsilon^2. \label{EqBenchmarkNoiseVar}
\end{flalign}
The benchmark fixes $X_{k-1}$ to a Gaussian family across all configurations and
sweeps high-likelihood state locations near the mean-reversion point $c$. It
tests Gaussian process noise and two non-Gaussian alternatives with matched
innovation variance. The benchmark uses $c=0.95$, $\alpha=0.1$, and
$\beta=0.005$. It propagates uncertainty through one transition step and
compares the resulting law against a Monte Carlo ground truth using the
Wasserstein distance. The benchmark evaluates 84 distinct configurations. The
UxHw configurations use the Autocorrelation subsystem because $X_{k-1}$ appears
multiple times in Equation~\ref{EqCubicRevertingProcessTransformed}.

\begin{table}[t]
    \caption[Cubic mean-reverting transition
    model benchmark results.]{UxHw configuration benchmarks against corresponding equal-accuracy
    and equal-speed Monte Carlo simulation for the \textbf{cubic mean-reverting}
    transition model (Equation~\ref{EqCubicRevertingProcess}). The benchmark
    uses the Autocorrelation subsystem.}
    \label{TableCubicRevertingEMCC}
    \centering
    \footnotesize
    \begin{tabular}{l r r|r r r|r r r}
    \toprule
    \multicolumn{3}{c|}{\textbf{UxHw}}         & \multicolumn{3}{c|}{\textbf{Iso-quality Monte Carlo}} & \multicolumn{3}{c}{\textbf{Iso-time Monte Carlo}} \\
    \midrule
        & Avg\textsuperscript{\textdagger}& Runtime      & Avg Iter. & Runtime       & Speedup Avg         & Att. Iter.              & Avg  & \%$W_1$ change \\
    Method               &     $W_1$      & ($\mu$s)     & Count     & Avg ($\mu$s)  & (MC$\to$UxHw)       & $\lceil$Count$\rceil$   &   $W_1$  & (MC$\to$UxHw) \\
    \midrule
    \rowcolor{a} UxHw 8  & 0.004554       & 814       & 2329.42   & 199341    &  \texttimes 192.36  & 13     & 0.047317 & -90.38\% \\
    \rowcolor{b} UxHw 16 & 0.001514       & 3708      & 20146.60  & 1724049   &  \texttimes 466.24  & 44     & 0.026404 & -94.27\% \\
    \rowcolor{a} UxHw 32 & 0.000551       & 16124     & 137254.45 & 11745577  &  \texttimes 715.84  & 192    & 0.012744 & -95.68\% \\
    \midrule
                         &                &              & \multicolumn{3}{c|}{\textit{(90-th percentile values)}} &\multicolumn{3}{c}{\textit{(90-th percentile values)}} \\
    \cmidrule{4-9}
    \rowcolor{a}\multicolumn{3}{>{\cellcolor{white}}l|}{}& 4056.72   & 347155     &  \texttimes 334.99   &    & 0.071936 & -93.67\% \\
    \rowcolor{b}\multicolumn{3}{>{\cellcolor{white}}l|}{}& 34930.35  & 2989172    &  \texttimes 808.37   &    & 0.040429 & -96.26\% \\
    \rowcolor{a}\multicolumn{3}{>{\cellcolor{white}}l|}{}& 238171.64 & 20381586   &  \texttimes 1242.17  &    & 0.019462 & -97.17\% \\
    \bottomrule
    \multicolumn{9}{p{0.9\textwidth}}{Avg: average. Iter: iteration. Att:
    attainable. \textsuperscript{\textdagger}UxHw is deterministic and the same
    inputs always yield the same output; the column represents the average
    across benchmark inputs.}
    \end{tabular}
\end{table}

Table~\ref{TableCubicRevertingEMCC} shows the results for this cubic
mean-reverting transition model benchmark. UxHw~8 achieves an average \texttimes
192.36 speedup against the corresponding iso-quality Monte Carlo simulations,
and as much as \texttimes 334.99 in the worst-performing 10\% of the MC batches.
UxHw~16 and UxHw~32 achieve \texttimes 466.24 and \texttimes 715.84 speedup on
average, and as much as \texttimes 808.37 and \texttimes 1242.17 in 10\% of the
benchmark cases where Monte Carlo performs worst. The speedup grows with
representation size, which matches the higher resolution available for tracking
the nonlinear cubic drift and its induced dependence structure under the
Autocorrelation subsystem.

The comparison against the iso-time Monte Carlo simulations for each UxHw
configuration follows a pattern similar with that of the speedup above. UxHw~8
achieves on average -90.38\% Wasserstein distance in the same latency
as the iso-time Monte Carlo simulation, and by as much as -93.67\% for the
worst-performing 10\% of the MC batches. Similarly, UxHw~16 and UxHw~32 achieve
-94.27\% and -95.68\% Wasserstein distances to the ground truth on average,
and by as much as -96.26\% and -97.17\%.

These results show that processor-native uncertainty tracking on UxHw yields
better quality distributions than the Monte Carlo alternative within the same
time constraint.

\subsubsection{Exponential observation model benchmark}

Exponential laws arise whenever a quantity changes at a rate proportional to its
current value. Such models show up often in radiation intensity modelling,
electrical circuits, and medical devices~\cite{smal2008particle}. One example is
the Beer--Lambert law which is frequently employed in pulse oximetry observation
models. 

Let $Y_k$ denote the random variable corresponding to $y_k$, $M_k$ to $m_k$, and
$H_k$ to $h_k$. Equation~\ref{EqBeerLambertObservationRandomVars} gives the
transformation this benchmark tests:
\begin{equation}
    Y_k = e^{-X_k}(1 + M_k) + H_k. \label{EqBeerLambertObservationRandomVars}
\end{equation}
This benchmark treats $H_k$ as iid additive noise, even though $h_k$ represents
heteroscedastic noise in Equation~\ref{EqBeerLambertObservationModel} and in the
corresponding state estimation filter benchmarks.

This benchmark reuses the input specification of
Equations~\ref{EqBenchmarkStateInit}--\ref{EqBenchmarkNoiseVar}, while testing
for single scale $s_x$:
\begin{flalign}
    X_k &\sim \mathcal{N}(\lambda_x, s_x), \\
    \lambda_x &\in \{0.55,\,0.8,\,0.9,\,0.95,\,1.0,\,1.1,\,1.45\}, \text{ and}\\
    s_x &= 0.1. 
\end{flalign}
It then tests three noise families for each of the multiplicative term $M_k$ and
the additive term $H_k$, with matched variance across families. Let $\sigma_m$
denote the target standard deviation for $M_k$, and let $\sigma_h$ denote the
target standard deviation for $H_k$, with $(\sigma_m,\sigma_h)=(0.05,\,0.01)$.
Equations~\ref{EqExponentialObsBenchmarkSpaceStart}--\ref{EqExponentialObsBenchmarkSpaceEnd}
summarise the benchmark configuration space:
\begin{flalign}
    M_k &\sim \mu_M(0,\sigma_m),\qquad \mu_M \in \{\mathcal{N},\,\mathcal{L}(b_m),\,\mathcal{U}(w_m)\}, \label{EqExponentialObsBenchmarkSpaceStart} \\
    H_k &\sim \mu_H(0,\sigma_h),\qquad \mu_H \in \{\mathcal{N},\,\mathcal{L}(b_h),\,\mathcal{U}(w_h)\}, \text{ with}\\
    b_m &= \sigma_m/\sqrt{2},\quad b_h = \sigma_h/\sqrt{2}, \text{ and}\\
    w_m &= 2\sqrt{3}\,\sigma_m,\quad w_h = 2\sqrt{3}\,\sigma_h. \label{EqExponentialObsBenchmarkSpaceEnd}
\end{flalign}
The benchmark propagates uncertainty through application of
Equation~\ref{EqBeerLambertObservationRandomVars} and compares the resulting law
against a Monte Carlo ground truth using the Wasserstein distance. The
benchmark evaluates 63 distinct configurations.

\begin{table}[t]
    \caption[Exponential observation model benchmark results.]{UxHw
    configuration benchmarks against corresponding equal-accuracy and
    equal-speed Monte Carlo simulation for the simplified \textbf{exponential}
    observation model (Beer--Lambert) with heteroscedastic uncertainty
    (Equation~\ref{EqBeerLambertObservationModel}). The UxHw configurations use
    the Autocorrelation subsystem.}
    \label{TableBeerLambertEMCC}
    \centering
    \footnotesize
    \begin{tabular}{l r r|r r r|r r r}
    \toprule
    \multicolumn{3}{c|}{\textbf{UxHw}}         & \multicolumn{3}{c|}{\textbf{Iso-quality Monte Carlo}} & \multicolumn{3}{c}{\textbf{Iso-time Monte Carlo}} \\
    \midrule
        & Avg\textsuperscript{\textdagger}& Runtime      & Avg Iter. & Runtime       & Speedup Avg         & Att. Iter.              & Avg  & \%$W_1$ change \\
    Method               &     $W_1$      & ($\mu$s)     & Count     & Avg ($\mu$s)  & (MC$\to$UxHw)       & $\lceil$Count$\rceil$   &   $W_1$  & (MC$\to$UxHw) \\
    \midrule
    \rowcolor{a} UxHw 8  & 0.001572       & 16603      & 1765.76   & 415282    &  \texttimes 25.01  & 71     & 0.006897 & -77.21\% \\
    \rowcolor{b} UxHw 16 & 0.000539       & 72614      & 13836.73  & 3254205   &  \texttimes 44.81  & 309    & 0.003361 & -83.96\% \\
    \rowcolor{a} UxHw 32 & 0.000219       & 326058     & 84928.40  & 19973971  &  \texttimes 61.26  & 1387   & 0.001575 & -86.10\% \\
    \midrule
                         &                &              & \multicolumn{3}{c|}{\textit{(90-th percentile values)}} &\multicolumn{3}{c}{\textit{(90-th percentile values)}} \\
    \cmidrule{4-9}
    \rowcolor{a}\multicolumn{3}{>{\cellcolor{white}}l|}{}& 3052.75   & 717964     &  \texttimes 43.24   &    & 0.010728 & -85.35\% \\
    \rowcolor{b}\multicolumn{3}{>{\cellcolor{white}}l|}{}& 23892.33  & 5619142    &  \texttimes 77.38   &    & 0.005302 & -89.83\% \\
    \rowcolor{a}\multicolumn{3}{>{\cellcolor{white}}l|}{}& 146651.76 & 34490441   &  \texttimes 105.78  &    & 0.002463 & -91.11\% \\
    \bottomrule
    \multicolumn{9}{p{0.9\textwidth}}{Avg: average. Iter: iteration. Att:
    attainable. \textsuperscript{\textdagger}UxHw is deterministic and the same
    inputs always yield the same output; the column represents the average
    across benchmark inputs.}
    \end{tabular}
\end{table}

Table~\ref{TableCubicRevertingEMCC} lists the benchmark results of uncertainty
tracking for this simplified Beer--Lambert model. The largest tested UxHw~32
configuration delivers the highest average speedup (\texttimes 61.26) compared
to the latency of the iso-quality Monte Carlo simulation. UxHw~16 achieves
\texttimes 44.81, while UxHw~8 achieves an average speedup of \texttimes 25.01.
In contrast with the cubic mean-reverting model of Section~\ref{SectionCubicRevertingBenchmark}, while UxHw achieves clear
speedups compared to the corresponding iso-quality Monte Caro simulations, the
UxHw delivers less efficient uncertainty tracking for this benchmark.

The iso-time Monte Carlo simulation in Table~\ref{TableBeerLambertEMCC} limits
the iteration count to match each UxHw configuration runtime. Compared to the
Wasserstein distance achieved by the corresponding iso-time Monte Carlo
simulations, UxHw~8, UxHw~16, and UxHw~32 achieve smaller $W_1$ on average:
-77.21\%, -83.96\%, and -86.10\% at matched latency, and as much as -85.35\%,
-89.83\%, and -91.11\% for the 10\% worst-performing MC batches.

The smaller speedups, and the larger iso-time accuracy gap, suggest that
the exponential observation map imposes a difficult uncertainty transform.
The model applies an exponential nonlinearity and composes two independent noise
sources. This structure increases the approximation burden on the finite UxHw
uncertainty representation and reduces the accuracy gained per unit of UxHw
compute, relative to the cubic mean-reverting benchmark, even though UxHw
remains more accurate than Monte Carlo at equal runtime.

\subsubsection{Exponential likelihood inference benchmark}

This section benchmarks the inference speed and accuracy of UxHw BayesLaplace
against importance sampling in particle filtering. The exponential observation
model of Equation~\ref{EqBeerLambertObservationModel} includes two noise terms,
so the importance-sampling baseline must marginalise over latent observation
noise when it evaluates particle weights. For this model, that
marginalisation is stable enough that the baseline remains well-behaved.

\begin{table}[tb]
    \centering
    \caption[Exponential likelihood model inference benchmark latency and
    accuracy results.]{Exponential likelihood model inference benchmark latency
    and accuracy results for importance sampling (MC) and BayesLaplace (UxHw).}
    \label{TableBeerLambertInferenceBenchmarkResults}
    \centering
    \footnotesize
    \begin{tabular}{lr|S[table-format=8.0,group-minimum-digits=4]S[table-format=1.6]S[table-format=1.6]}
    \toprule
    Method & Fidelity & {Latency ($\mu$s) \scalebox{0.8}{\rotatebox{90}{\ding{222}}}} & {Mean $W_1$} & {$W_1$ Std} \\
    \midrule
    \rowcolor{a} MC   & 10   & 6875     & 0.026141 & 0.010015 \\
    \rowcolor{b} MC   & 20   & 13856    & 0.019122 & 0.007059 \\
    \rowcolor{a} MC   & 50   & 35254    & 0.012480 & 0.004477 \\
    \rowcolor{b} MC   & 100  & 70714    & 0.008948 & 0.003143 \\
    \rowcolor{a} UxHw & 8    & 89663    & 0.002206 & 0.001898 \\
    \rowcolor{b} MC   & 500  & 367572   & 0.004123 & 0.001430 \\
    \rowcolor{a} UxHw & 16   & 659346   & 0.001489 & 0.000487 \\
    \rowcolor{b} MC   & 1000 & 739641   & 0.002991 & 0.001035 \\
    \rowcolor{a} UxHw & 32   & 20257900 & 0.001329 & 0.000529 \\
    \bottomrule
    \end{tabular}
\end{table}

\begin{figure}[tb]
    \centering
    \includegraphics[trim={0cm 0cm 0cm 0.46cm},clip,width=0.95\textwidth]{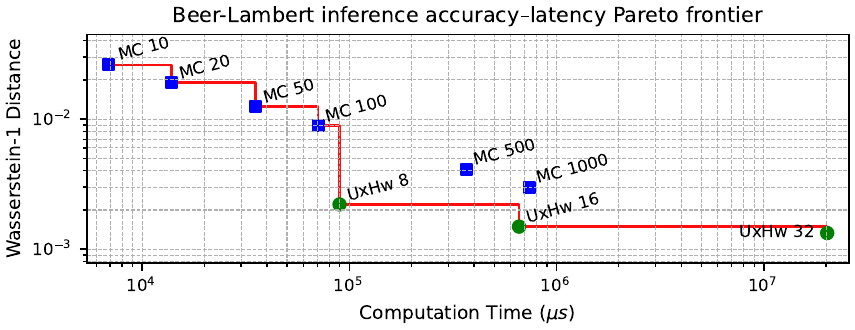}
    \caption[Pareto frontier plot for accuracy and latency of the inference
    mechanism for the exponential likelihood.]{Pareto frontier plot for the
    distributional accuracy and latency of the inference mechanism for the
    exponential likelihood system. Green circles are UxHw BayesLaplace inference
    and blue squares are inference via importance
    sampling.\label{FigBeerLambertAccuracyLatencyPareto}}
\end{figure}

For each inference instance, the benchmark computes the posterior law using UxHw
BayesLaplace on UxHw~8, UxHw~16, and UxHw~32, and using importance sampling with
particle count $n \in \{10, 20, 50, 100, 500, 1000\}$.
Table~\ref{TableBeerLambertInferenceBenchmarkResults} lists the mean and
standard deviation of the Wasserstein distance against a Monte Carlo
ground-truth posterior together with the measured latency, and
Figure~\ref{FigBeerLambertAccuracyLatencyPareto} shows the corresponding
accuracy--latency Pareto frontier.

UxHw~8 attains Wasserstein distance \num{0.002206} with
\qty{89663}{\micro\second} latency, which is more accurate than all tested
importance-sampling configurations, including 1000 particles, which reaches
distance \num{0.002991} at \qty{739641}{\micro\second}. UxHw~16 reduces the mean
distance further to \num{0.001489} at \qty{659346}{\micro\second}. UxHw~32
improves the mean again to \num{0.001329}, but raises latency to
\qty{20257900}{\micro\second}. For this model, larger UxHw configurations
therefore give smaller gains in accuracy than the extra runtime they require.

The importance-sampling results improve steadily as the particle count
increases, but remain behind UxHw in accuracy across the tested range. The
Pareto frontier in Figure~\ref{FigBeerLambertAccuracyLatencyPareto} therefore
favours UxHw~8 and UxHw~16 over the tested importance-sampling baselines. These
results show that, for this benchmark, UxHw BayesLaplace gives a better
accuracy--latency trade-off than importance sampling for the latency regimes of
\qty{100}{\milli\second} to \qty{20}{\second}. The Monte Carlo alternative
remains a flexible alternative able to address lower latency requirements at the
expense of lower distributional accuracy.

\section{Summary of results}

This paper presents a Bayesian filtering technique that uses processor-native
uncertainty tracking for both uncertainty propagation and inference, and
compares it to particle-based methods via the particle filter. The
technique uses UxHw to construct approximate encoded laws for forward
uncertainty propagation, and it implements inference via deterministic
hierarchical importance restructuring. In this way, UxHw uses deterministic
distributional arithmetic to compute model outputs and Bayesian posteriors with
deterministic latency and memory for arbitrary models written as program code,
without relying on extra random sampling. 

Across the three common nonlinear systems in this paper, the UxHw approach
yields competitive estimation RMSE against particle filters while also
maintaining smaller latencies when noise terms follow Gaussian distributions.
The UxHw-based filters are deterministic, so they produce the same result for
the same input data. In tests involving inputs with both Gaussian and
non-Gaussian distributions, for the same distribution accuracy measured by
Wasserstein distance, the UxHw approach consistently outperforms the
Monte-Carlo-based estimator of the particle filters. In the average Monte Carlo
cases, it achieves at least \texttimes24.73 speedup against the iso-quality
Monte Carlo in the stochastic volatility observation model, and as much as
\texttimes805.18 for UxHw~8 in the stochastic volatility transition model. For
the worst-performing 10\% of Monte Carlo batches, the UxHw approach achieves
speedup as high as \texttimes1424.58 against the iso-quality Monte Carlo, for
UxHw~32 in the Gordon--Salmond--Smith transition model.

The UxHw approach also achieves significant $W_1$-accuracy improvements over
Monte Carlo in the same latency budget. In the average Monte Carlo cases, UxHw
achieves at least -76.69\% Wasserstein distance, for UxHw~8 in the
stochastic volatility observation model, and as much as -96.10\% for UxHw~32 in
the Gordon--Salmond--Smith transition model. For the worst-performing 10\% of
the Monte Carlo batches, UxHw achieves as much as -97.95\% Wasserstein
distance against the iso-time Monte Carlo simulation.

For posterior inference computation, the Pareto-frontier plots show that UxHw
consistently dominates Monte Carlo in both $W_1$ accuracy and computation
latency for the three observation models in this paper. In the stochastic
volatility and Gordon--Salmond--Smith systems, UxHw~8 dominates Monte Carlo
with 50 samples and with 100 samples. In the Beer--Lambert system, Monte Carlo
is faster per sample, so UxHw~8 dominates MC 500 and MC 1000.

The results also show that the scaling of the UxHw approach depends on model
structure. The stochastic volatility transition model and the
Gordon--Salmond--Smith observation model yield comparatively worse results for
higher UxHw $\eta$ than for UxHw~8, while still remaining better than the Monte
Carlo alternative. The other four models all improve progressively with higher
UxHw fidelity. The performance results are also not smooth across models. While
some models yield UxHw speedup in the hundreds, the stochastic volatility
observation and exponential observation models yield speedup in the decades,
which suggests degraded UxHw performance in these configurations, while still
remaining better than Monte Carlo.

UxHw representations with $\eta=16$ and $\eta=32$ remain performant and yield
results competitive with, or better than, the Monte Carlo alternative. However,
the speed of UxHw~8 makes it the most attractive choice for dynamic online
Bayesian filtering in resource-constrained systems. In systems with greater
processing or memory capacity, engineers can choose a higher UxHw representation
size and trade latency for distributional accuracy. This trade-off becomes
important in applications that depend on tail accuracy in the estimate
distribution.

\printbibliography[heading=bibintoc, title={References}]
\end{document}